\documentclass[opre,nonblindrev,nameyear]{informs3_hide}
\OneAndAHalfSpacedXI
\usepackage{amsmath,amssymb,bm,graphicx,booktabs,array}
\usepackage{xcolor}

\usepackage{algorithm,algorithmicx}
\usepackage[noend]{algpseudocode}
\algdef{SE}[SUBALG]{Indent}{EndIndent}{}{\algorithmicend\ }
\algtext*{Indent}
\algtext*{EndIndent}

\usepackage[round,authoryear]{natbib}
\usepackage{xurl}
\usepackage[hidelinks]{hyperref}
\usepackage[capitalize,noabbrev]{cleveref}
\TheoremsNumberedThrough
\EquationsNumberedThrough
\DeclareFontShape{OT1}{cmr}{bx}{sc}{<->ssub*cmr/bx/n}{}

\newcommand{\R}{\mathbb R}
\newcommand{\E}{\mathbb E}
\newcommand{\Var}{\operatorname{Var}}
\newcommand{\Prob}{\mathbb P}
\newcommand{\cF}{\mathcal F}
\newcommand{\cE}{\mathcal E}
\newcommand{\cO}{\mathcal O}
\newcommand{\norm}[1]{\lVert #1\rVert}
\newcommand{\ip}[2]{\langle #1,#2\rangle}
\newcommand{\QME}{\operatorname{QME}}
\crefname{assumption}{Assumption}{Assumptions}
\Crefname{assumption}{Assumption}{Assumptions}

\RUNAUTHOR{Cao et al.}
\RUNTITLE{Quantum Sampling in Gradient-Based Simulation Optimization}
\TITLE{Beyond Classical Minimax Rates: Quantum Sampling in Gradient-Based Simulation Optimization}
\ARTICLEAUTHORS{
\AUTHOR{Hao Cao, Jian-Qiang Hu, Mingjie Hu, Qixin Wang}
\AFF{School of Management, Fudan University,
Shanghai 200433, China,
\EMAIL{hcao21@m.fudan.edu.cn},
\EMAIL{hujq@fudan.edu.cn},
\EMAIL{23110690009@m.fudan.edu.cn},
\EMAIL{23110690014@m.fudan.edu.cn}}
}

\allowdisplaybreaks[4]

\newcommand{\bx}{\mathbf{x}}
\newcommand{\bg}{\mathbf{g}}
\newcommand{\bw}{\mathbf{w}}
\newcommand{\bs}{\mathbf{s}}
\newcommand{\bunit}{\mathbf{e}}
\newcommand{\bM}{\mathbf{M}}
\newcommand{\bP}{\mathbf{P}}
\newcommand{\bI}{\mathbf{I}}
\newcommand{\bone}{\mathbf{1}}
\newcommand{\bzero}{\mathbf{0}}

\ABSTRACT{We study gradient-based search for a class of convex simulation optimization problems with gradients estimated from noisy outputs. With independent outputs, the optimal classical convergence rate for the expected objective gap is of order \(\cO(N^{-2/3})\), where \(N\) is the query budget. 
We apply quantum mean estimation to paired finite differences and progressively increase their estimation accuracy as the perturbations shrink, improving the convergence rate to \(\cO(N^{-4/3})\). 
Furthermore, we show that coupling nearby evaluations improves the classical and quantum convergence rates to \(\cO(N^{-2/(3-\kappa)})\) and \(\cO(N^{-4/(3-\kappa)})\), respectively, where \(\kappa\in[0,1]\) describes the variance reduction in paired outputs. Thus, quantum estimation halves the accuracy exponent in query complexity across coupling orders. 
Moreover, we establish matching lower bounds and show that these accuracy exponents are minimax optimal for fixed decision dimension, even when query locations and batch sizes are adaptive. Reconstruction from feasible differences retains the exponents when every simulation query must be feasible and accommodates boundary optima.}
\KEYWORDS{gradient-based simulation optimization; stochastic approximation; simultaneous perturbation; query complexity; quantum sampling}
\begin{document}
\maketitle

\section{Introduction}
\label{sec:introduction}
Gradient-based algorithms are widely used to optimize the expected performance of stochastic systems with continuous decisions \citep[cf., e.g.,][]{fu2002, fu2008you, fu2012conditional}. Even when the expected objective and its gradient are unavailable analytically, gradients can be estimated from noisy simulation outputs \citep{golovin2017google,caoblackbox2025,huquantile2025}. Component-wise finite differences perturb each decision variable separately \citep{kiefer1952}, whereas simultaneous perturbation stochastic approximation (SPSA) perturbs all variables simultaneously and uses outputs at only two locations \citep{spall1992}. In both cases, shrinking the perturbations reduces derivative bias but can increase the variance of the gradient estimate.

Quantum mean estimation offers the possibility of improving the convergence order of simulation optimization beyond the gains from classical variance reduction \citep[cf., e.g.,][]{kleinman1999, he2024adaptive}.
For a scalar output with standard deviation at most \(s\), classical averaging attains a root mean squared error (RMSE) tolerance \(\tau>0\) using \(\cO(1+s^2/\tau^2)\) samples. With coherent access to a reversible sampler, quantum mean estimation attains the same tolerance using \(\cO(1+s/\tau)\) queries \citep{kothari2023}.

However, the quadratic speedup in mean estimation does not automatically carry over to optimization. A quantum sampler encodes possible difference quotients in superposition, but directly measuring its output yields only one noisy realization. The speedup comes from coherent queries that estimate the mean to a prescribed accuracy at a fixed decision and perturbation. An essential challenge remains in choosing the estimation accuracy, which affects both the quantum query cost of each update and the number of updates needed to attain a target expected objective gap. High precision early in the search can consume queries with little benefit to the final solution, while insufficient precision later can impede convergence. Shrinking perturbations further complicates this allocation because it reduces finite-difference bias but may increase the variance of the difference quotients and hence their estimation cost. Obtaining a quadratic improvement in total optimization query complexity requires choosing estimation tolerances jointly with perturbation sizes and controlling how the resulting errors affect subsequent decisions.

To address this challenge, we develop quantum acceleration with coupled evaluations (QACE), a gradient-based algorithm that achieves a quadratic improvement in total query complexity by coordinating mean-estimation accuracy with shrinking perturbations.
At each iteration, we form a finite difference from paired outputs and estimate its conditional mean. Our constant-step analysis shows that the contribution of early estimation errors to the objective-gap bound decreases over subsequent updates. This allows us to use coarse estimates early in the search and progressively tighten the estimation tolerance as the perturbations shrink. The resulting schedule yields a geometrically decreasing bound on the expected objective gap and geometrically increasing query bounds per update. Summing these query bounds over all updates preserves the quadratic improvement in the accuracy dependence of total query complexity across coupling orders.

\subsection{Related Work}
Stochastic approximation (SA) provides recursive gradient-based methods for a broad range of objectives estimated by simulation \citep{kushner2003sa, borkar2009stochastic}. SA methods have been developed for quantile optimization \citep{huquantile2022, huquantile2025}, conditional value-at-risk optimization \citep{hucvar2025, wangcvar2025}, optimization of distortion risk measures \citep{jiang2025stochastic,cao2026estimation}, likelihood-free parameter estimation \citep{li2026beyond}, and analysis of dynamic systems \citep{jiang2023quantile,cao2023infinitesimal,chen2026online,li2026convergence,che2026stochastic,wang2026long}. Recently, \citet{wangzhang2026llmmas} apply randomized two-point perturbations for optimization in large language model-powered multi-agent simulations.
Common random numbers (CRN) can reduce the variance of paired differences and improve the convergence rate of SA \citep{kleinman1999}. 
\citet{he2024adaptive} develop adaptive importance sampling for SA, but their method requires likelihood ratios that are unavailable from our noisy-output oracle.

To determine the benefit of quantum estimation, convergence bounds across iterations must be linked to the simulation effort required by each update. \citet{hufu2025} derive finite-time error bounds that allow finite-difference bias and noise growing as the perturbations shrink. Their bounds describe convergence for given bias and noise sequences, but do not determine how to allocate the queries needed for quantum estimation. We jointly choose estimation tolerances and perturbation sizes to preserve geometric convergence, and account for the increasing query cost over the entire search.

Establishing that this improvement is optimal further requires lower bounds that account for coupling and coherent quantum access. For independent noisy function evaluations, \citet{yu2024} characterize the classical minimax accuracy exponent using batched estimates from function values. Their lower bound concerns independent classical observations, so it neither constrains coherent quantum queries nor determines the optimal accuracy exponent when paired outputs are coupled. We construct paired simulators that yield matching classical and quantum lower bounds for the accuracy exponents under the same coupling condition. These bounds establish an optimal quadratic separation even when classical methods exploit coupling and adapt their query locations and sampling effort.

Quantum optimization results based on gradient or uniformly accurate function oracles do not directly determine the cost of optimization from noisy simulation outputs. The stochastic gradient methods of \citet{sidford2023} assume unbiased gradient estimates. In contrast, finite differences from simulation outputs introduce bias and may have increasing variance as the perturbations shrink. Related zeroth-order methods assume access to uniformly accurate objective approximations \citep{kim2025,gong2025}. A bound on the variance of paired differences does not provide the uniform approximation error these methods require, and obtaining the prescribed accuracy itself consumes simulation queries. 
To establish query complexity bounds under noisy-output access, we analyze paired simulation differences and account for their estimation cost over all updates.

Discontinuous simulation responses create a further obstacle because a smooth expected objective need not have Lipschitz sample responses or unbiased sample gradients. 
The CRN convergence analysis of \citet{kleinman1999} assumes bounded sample-path derivatives, which are not required in our setting. 
\citet{liu2024} apply quantum mean estimation to two-point differences with shared inputs under sample-wise Lipschitz continuity. This condition keeps the difference quotients bounded as the perturbations shrink. 
However, for discontinuous event outputs, the variance of the difference quotients can increase as the perturbations shrink.
The zeroth-order gradient-estimation results of \citet{ozgul2026} assume that sample-path gradients are unbiased and have uniformly bounded variance.
But for discontinuous event responses, sample-path gradients can be zero almost surely (a.s.) even when the expected objective has a nonzero gradient.
We work directly with the variance of paired difference quotients and coordinate its growth with finite-difference bias and estimation accuracy. This allows us to establish a quadratic improvement in total query complexity across coupling orders, including simulations with discontinuous outputs.

When simulations are available only at feasible decisions, projecting the iterates does not ensure that finite-difference queries remain feasible. 
\citet{lam2025} address this issue on probability simplices using Dirichlet perturbations. To accommodate more general convex feasible sets, we reconstruct gradients from differences along feasible segments. Our approach allows lower-dimensional feasible sets and boundary optima while preserving the quadratic quantum speedup in query complexity.

\subsection{Main Contributions}
Our main contribution is to establish a quadratic improvement in the accuracy dependence of optimization query complexity across the full range of coupling orders. With independent outputs, classical methods attain an expected objective gap of order \(\cO(N^{-2/3})\) after \(N\) queries, and QACE improves this rate to \(\cO(N^{-4/3})\).

This quadratic improvement persists even when classical estimation already benefits from strong coupling. We characterize the variance reduction shared by both methods through a coupling order \(\kappa\in[0,1]\), with \(\kappa=0\) covering independent outputs and larger values describing stronger variance reduction.
As a special case, \(\kappa=1\) covers CRN couplings with a quadratic variance bound on paired differences.
To attain an expected objective gap of at most \(\varepsilon\), classical paired averaging with centered perturbations uses \(\cO(d^2\varepsilon^{-(3-\kappa)/2})\) queries, where \(d\) is the decision dimension.
QACE reduces this bound to \(\cO(d^{3/2}\varepsilon^{-(3-\kappa)/4})\). The same \(\kappa\) governs both bounds, and quantum mean estimation halves the power of \(\varepsilon^{-1}\) across the full range \(\kappa\in[0,1]\). 
Equivalently, these bounds yield the objective-gap rates \(\cO(N^{-2/(3-\kappa)})\) and \(\cO(N^{-4/(3-\kappa)})\) for fixed \(d\), respectively.

Our second contribution is to prove minimax optimality of these accuracy exponents for fixed decision dimension, showing that adaptive classical sampling cannot recover the quantum improvement. We prove matching classical and quantum lower bounds under the same coupling condition, allowing adaptive query locations and batch sizes. Table~\ref{tab:comparison} summarizes the bounds.

Our third contribution is to preserve the same quadratic improvement when every simulation query must be feasible, including at boundary optima. We obtain this guarantee by reconstructing gradients from feasible differences. We also establish convergence with diminishing step sizes and guarantees under a deterministic query budget.

\begin{table}[htb]
\TABLE{
Simulation-query bounds for a target expected objective gap \(\varepsilon>0\) under the same coupling order \(\kappa\).  \label{tab:comparison}}{
\begin{tabular*}{0.9\linewidth}{@{\extracolsep{\fill}}lcc@{}}
\toprule
Method & Upper bound & Lower bound \\
\midrule
Classical paired averaging & \(\cO(d^2\varepsilon^{-(3-\kappa)/2})\) & \(\Omega(d^{(1+\kappa)/2}\varepsilon^{-(3-\kappa)/2})\) \\
\addlinespace
QACE & \(\cO(d^{3/2}\varepsilon^{-(3-\kappa)/4})\) & \(\Omega(d^{(3+\kappa)/4}\varepsilon^{-(3-\kappa)/4})\) \\
\bottomrule
\end{tabular*}}{}
\end{table}

The rest of this paper is organized as follows. Section~\ref{sec:problem} introduces the optimization problem and quantum mean estimation, and develops the gradient-based algorithms. Sections~\ref{sec:theory} and \ref{sec:lower} establish the convergence and query complexity bounds. Section~\ref{sec:numerical} presents numerical results, and Section~\ref{sec:conclusion} concludes. 

\section{Gradient-Based Optimization and Quantum Sampling}
\label{sec:problem}
We focus on the simulation optimization problem \(\min_{\bx\in\mathcal X}f(\bx)\), where \(f(\bx)\mathrel{:=}\E[Y(\bx,\xi)]\) denotes the objective function and \(\mathcal X\subseteq\R^d\) denotes a nonempty, closed, and convex feasible region.  
\(\bx\) denotes the decision, \(\xi\) contains the uncertainty in one simulation experiment, and we assume \(\E[|Y(\bx,\xi)|]<\infty\).
For notational simplicity, we write \(\bx_*\mathrel{:=}\operatorname{argmin}_{\bx\in\mathcal X}f(\bx)\) and \(f_*\mathrel{:=}f(\bx_*)\).

Gradient-based search uses the local sensitivity of \(f\) to determine how to improve the current decision. When the gradient is unavailable analytically, we consider the update 
\[
\bx_{k+1}=\Pi_{\mathcal X}(\bx_k-\eta_k\bg_k),
\quad k=0,1,\ldots,
\]
where \(\bg_k\) denotes the estimate of \(\nabla f(\bx_k)\), \(\eta_k>0\) is the step size, and \(\Pi_{\mathcal X}\) denotes the Euclidean projection that prevents the iterate from leaving the feasible set \(\mathcal X\). 
The main task is to construct \(\bg_k\) from simulation outputs. Finite differences approximate the gradient using the difference \(f(\bx)-f(\bx')\) at nearby decisions \(\bx\) and \(\bx'\), so constructing \(\bg_k\) requires estimating the means of paired simulation differences, whose query cost can be reduced through quantum sampling.

\subsection{Quantum Mean Estimation}
\label{app:qvr}
To estimate the difference \(f(\bx)-f(\bx')\), we generate paired outputs \(Y(\bx,\xi)\) and \(Y(\bx',\xi')\) with their prescribed marginal distributions, so that \(\E[Y(\bx,\xi)-Y(\bx',\xi')]=f(\bx)-f(\bx')\). Coupling the inputs can reduce the variance of this difference. For example, CRN imposes \(\xi'=\xi\). 
Even with coupling, classical averaging may require many queries to attain a small estimation error.
To reduce this cost, we use quantum mean estimation for the same paired difference.

Specifically, quantum mean estimation takes a reversible scalar sampler \(\mathcal A\) with output \(Y\), a constant \(s\ge0\) bounding its standard deviation, and an RMSE tolerance \(\tau>0\). We assume coherent access to the reversible sampler \(\mathcal A\), its inverse, and controlled queries. 
We write \(\QME[\mathcal A,s,\tau]\) for the resulting estimate of the sampler's output mean. Classical averaging attains the same tolerance using \(\max\{1,\lceil s^2/\tau^2\rceil\}\) independent outputs.
We write \(\mathsf q(\widehat\mu)\) for the number of sampler queries used to produce \(\widehat\mu\). Lemma~\ref{lem:qvr} below provides the RMSE control needed for gradient estimation. Its proof converts a mean estimate with constant success probability into an RMSE guarantee and uses randomized telescoping for unbiasedness \citep{kothari2023, rhee2015, sidford2023}.
We then use these quantum mean estimates to construct gradient updates in Sections~\ref{sec:methodsa} and \ref{sec:methodsb}.

\begin{lemma}
\label{lem:qvr}
There exists an estimate \(\widehat\mu\) satisfying \(\E[(\widehat\mu-\E[Y])^2]\le\tau^2\). The unbiased implementation satisfies \(\E[\widehat\mu]=\E[Y]\) and \(\E[\mathsf q(\widehat\mu)]\le\mathcal C_1(1+s/\tau)\), for some constant \(\mathcal C_1>0\). The implementation with a deterministic query cap satisfies \(\mathsf q(\widehat\mu)\le \mathcal C_1(1+s/\tau)\) a.s.
\end{lemma}


\subsection{Centered Perturbations}
\label{sec:methodsa}
To estimate all gradient components using only two simulation locations, we use simultaneous perturbations. We first consider the setting where simulation queries may lie outside \(\mathcal X\), with the simulator defined at the perturbed locations. We assume that \(\mathcal X\) has nonempty interior and \(\bx_*\in\operatorname{int}\mathcal X\), as in \cite{hufu2025}. At \(\bx\in\mathcal X\), we draw a random direction \(\boldsymbol\Delta\mathrel{:=}(\Delta_1,\ldots,\Delta_d)^\top\), independently of \(\bx\) and past observations, with independent components satisfying \(\Prob(\Delta_j=1)=\Prob(\Delta_j=-1)=1/2\). We use an independent sequence \(\{\boldsymbol\Delta_k\}\) of these random directions, generated independently of the simulator randomization. The directional derivative along \(\boldsymbol\Delta/\sqrt d\) is approximated by the mean of the centered simulation difference,
\begin{equation}
\label{eq:centereddifference}
\begin{aligned}
 \mu_c(\bx,\boldsymbol\Delta)&\mathrel{:=}\E\Big[\frac{Y(\bx+c\boldsymbol\Delta/\sqrt d,\xi)-Y(\bx-c\boldsymbol\Delta/\sqrt d,\xi')}{2c}\,\Big|\,\bx,\boldsymbol\Delta\Big]\\
 &=\frac{f(\bx+c\boldsymbol\Delta/\sqrt d)-f(\bx-c\boldsymbol\Delta/\sqrt d)}{2c},
\end{aligned}
\end{equation}
where \(c\in(0,1]\) denotes the perturbation size and \(\bx\pm c\boldsymbol\Delta/\sqrt d\) are the perturbed simulation inputs. The independent signs ensure that \(\E[\sqrt d\mu_c(\bx,\boldsymbol\Delta)\boldsymbol\Delta]\) approximates \(\nabla f(\bx)\) as \(c\) decreases. To estimate \(\mu_c(\bx,\boldsymbol\Delta)\), let \(\mathcal A(\bx,\boldsymbol\Delta,c)\) prepare the random quotient inside the expectation in \eqref{eq:centereddifference}, using one paired simulator query and reversible arithmetic. 
Forming the difference before mean estimation retains the variance reduction from coupling. The estimate \(\widehat\mu=\QME[\mathcal A(\bx,\boldsymbol\Delta,c),s,\tau]\) has RMSE at most \(\tau\). 
In Algorithm~\ref{alg:qace} below, we illustrate how these estimates can be applied to update the decision and to solve the optimization problem.

\begin{algorithm}[htb]
\caption{QACE with centered perturbations.\protect\label{alg:qace}}
\begin{algorithmic}[1]
\State \textbf{Input} initial decision \(\bx_0\in\mathcal X\); 
step- and perturbation-size sequences \(\{\eta_k\}\) and \(\{c_k\}\);
estimation tolerances \(\{\tau_k\}\); 
standard-deviation bounds \(\{s_k\}\); 
random directions \(\{\boldsymbol\Delta_k\}\).
\State \textbf{Iterate} for \(k=1,2,\ldots,K\).
\Indent
\State \textbf{Mean estimation}
\begin{equation*}
 \widehat\mu_{k-1}=\QME[\mathcal A(\bx_{k-1},\boldsymbol\Delta_{k-1},c_{k-1}),s_{k-1},\tau_{k-1}].
\end{equation*}
\State \textbf{Gradient estimation and decision update}
\begin{equation}\label{eq:qaceupdate}
\bg_{k-1}=\sqrt d\widehat\mu_{k-1}\boldsymbol\Delta_{k-1} \quad\text{and}\quad \bx_k=\Pi_{\mathcal X}(\bx_{k-1}-\eta_{k-1}\bg_{k-1}).
\end{equation}
\EndIndent
\State \textbf{Return} \(\bx_K\).
\end{algorithmic}
\end{algorithm}

If we use one observed pair in place of mean estimation, we recover \(\bg_k=\sqrt d[Y(\bx_k+c_k\boldsymbol\Delta_k/\sqrt d,\xi)-Y(\bx_k-c_k\boldsymbol\Delta_k/\sqrt d,\xi')]\boldsymbol\Delta_k/(2c_k)\), i.e., the usual two-sided SPSA estimate. 
In Algorithm~\ref{alg:qace}, QACE instead estimates \(\mu_{c_k}(\bx_k,\boldsymbol\Delta_k)\) to tolerance \(\tau_k\), where decreasing \(\tau_k\) may increase the simulation effort per iteration.
Classical averaging requires growing batches of paired outputs, whereas quantum mean estimation can achieve the same tolerances with fewer coherent queries.
We show in Section~\ref{sec:constant} that the quadratic quantum advantage in mean estimation extends to the total query complexity after accounting for the increasing estimation cost over all updates.

\subsection{Feasible Reconstruction}
\label{sec:methodsb}
When the simulator can be queried only on \(\mathcal X\), centered perturbations in \eqref{eq:centereddifference} may be unavailable near the boundary. To keep every query feasible, we compare the current decision with points on feasible segments and reconstruct the gradient from these directional estimates. We take \(\mathcal X\) to be compact and allow lower-dimensional sets and boundary optima. We define \(\mathcal T\mathrel{:=}\operatorname{span}\{\bx-\bx':\bx,\bx'\in\mathcal X\}\), with \(\dim\mathcal T\ge1\), and let \(\bP\) be the orthogonal projection onto \(\mathcal T\). We write \(\nabla_{\mathcal T}f\) and \(\nabla_{\mathcal T}^2f\) for the gradient and Hessian along this space, respectively.

We first select feasible directions from which the gradient can be reconstructed. At \(\bx\), we choose anchors \(\bx^{[1]},\ldots,\bx^{[m]}\in\mathcal X\), with \(m\ge \dim\mathcal T\), and define \(\bw_i\mathrel{:=}\bx^{[i]}-\bx\) and \(\bM_{\bx}\mathrel{:=}\sum\nolimits_{i=1}^m\bw_i\bw_i^\top\). We require \(\bM_{\bx}\succeq\lambda\bP\) on \(\mathcal T\) for a uniform constant \(\lambda>0\). The reconstruction vectors \(\bs_i\mathrel{:=}\bM_{\bx}^\dagger\bw_i\), where \(\dagger\) denotes the Moore--Penrose inverse, then satisfy \(\sum\nolimits_i\bs_i\bw_i^\top=\bP\) and \(\sum\nolimits_i\bs_i\ip{\nabla_{\mathcal T}f(\bx)}{\bw_i}=\nabla_{\mathcal T}f(\bx)\). 
If the anchors are fixed and \(\operatorname{span}\{\bx^{[i]}-\bx^{[1]}: i=2,\ldots,m\}=\mathcal T\), then the condition \(\bM_{\bx}\succeq\lambda\bP\) is satisfied for some \(\lambda>0\) uniformly over \(\bx\in\mathcal X\).
As a concrete example, consider the probability simplex \(\mathcal S_d\mathrel{:=}\{\bx\in\R^d\mid\bx\ge\bzero,\ \bone^\top\bx=1\}\), where \(d\ge2\) and \(\bone\) denotes the vector of ones. We can use the coordinate unit vectors \(\bunit_i\) as anchors, which yields \(m=d\), \(\dim\mathcal T=d-1\), and \(\lambda=1\). 

We next estimate these directional derivatives with bias of order \(c^2\) using only feasible queries. For \(0<t\le1\), the one-sided expansion \([f(\bx+t\bw_i)-f(\bx)]/t=\ip{\nabla_{\mathcal T}f(\bx)}{\bw_i}+(t/2)\bw_i^\top\nabla_{\mathcal T}^2f(\bx)\bw_i+\cO(t^2)\) contains a term linear in \(t\), which we remove by Richardson extrapolation. For \(\bw_i\ne\bzero\), we let \(\mathcal A^F(\bx,\bw_i,c)\) prepare
\begin{equation}
\label{eq:richardson}
 \frac{4Y(\bx+c\bw_i/2,\xi_1)-4Y(\bx,\xi'_1)-Y(\bx+c\bw_i,\xi_2)+Y(\bx,\xi'_2)}{c},
\end{equation}
where the pairs \((\xi_1,\xi'_1)\) and \((\xi_2,\xi'_2)\) are independent, with coupling retained within each pair. This sampler uses two paired queries and reversible arithmetic. Using the unbiased implementation and a common standard-deviation bound \(s^F\) for these samplers, we independently compute \(\widehat\mu_i^F=\QME[\mathcal A^F(\bx,\bw_i,c),s^F,\tau\sqrt{\lambda/\dim\mathcal T}]\). These estimates satisfy \(\E[\widehat\mu_i^F]=[4f(\bx+c\bw_i/2)-f(\bx+c\bw_i)-3f(\bx)]/c\) and \(\E[\widehat\mu_i^F]-\ip{\nabla_{\mathcal T}f(\bx)}{\bw_i}=\cO(c^2)\). For capped estimates, we use the scalar tolerance \(\tau\sqrt{\lambda/m}\). If \(\bw_i=\bzero\), we set \(\widehat\mu_i^F=0\) without querying the simulator.

We obtain the feasible reconstruction update in \eqref{eq:feasibleupdate} by replacing the centered differences in Algorithm~\ref{alg:qace} with directional estimates along feasible segments:
\begin{equation}
\label{eq:feasibleupdate}
 \widetilde{\bg}_{k-1}=\sum\nolimits_{i=1}^m\bs_{k-1,i}\widehat\mu_{k-1,i}^F
 \quad\text{and}\quad \bx_k=\Pi_{\mathcal X}(\bx_{k-1}-\eta_{k-1}\widetilde{\bg}_{k-1}),
\end{equation}
where the subscript \(k-1\) within \(\widehat\mu_{k-1,i}^F\) indicates quantities formed at \(\bx_{k-1}\) with \(c_{k-1}\) and \(\tau_{k-1}\). Projection and convexity imply \(\bx_k\in\mathcal X\) and \(\bx_k+t\bw_{k,i}=(1-t)\bx_k+t\bx_k^{[i]}\in\mathcal X\) for every \(k\ge0\), \(i=1,\ldots,m\), and \(t\in\{0,c_k/2,c_k\}\). Thus, every simulation query is feasible, including queries made from boundary decisions.

\section{Convergence Rate Analysis}
\label{sec:theory}
We use the mean-estimation guarantee in Lemma~\ref{lem:qvr} to show how increasing precision preserves constant-step convergence and reduces the total query cost under the same coupling condition.
Throughout the paper, for positive scalar functions \(a\) and \(b\),
we use \(a=\cO(b)\) and \(a=\Omega(b)\) to denote \(a\le Cb\) and \(a\ge cb\), respectively, for some constants \(C,c>0\) independent of the varying parameters in the stated asymptotic regime.

\begin{assumption}
\label{ass:objective}
On an open convex neighborhood containing \(\mathcal X\) and all simulation inputs used by the procedure, we have \(f\in C^2\) and, for all \(\bx,\bx'\),
\begin{equation*}
\begin{aligned}
 f(\bx')&\ge f(\bx)+\ip{\nabla f(\bx)}{\bx'-\bx}+\frac\rho2\norm{\bx'-\bx}^2,\\
 \norm{\nabla f(\bx)-\nabla f(\bx')}&\le L_1\norm{\bx-\bx'},\quad\text{and}\quad
 \norm{\nabla^2f(\bx)-\nabla^2f(\bx')}\le L_2\norm{\bx-\bx'},
\end{aligned}
\end{equation*}
for constants \(0<\rho\le L_1<\infty\) and \(0<L_2<\infty\). 
For feasible reconstruction~\eqref{eq:feasibleupdate} on a lower-dimensional set, the neighborhood and derivatives are taken in its affine hull.
\end{assumption}

\begin{assumption}
\label{ass:pair}
At distinct simulation inputs \(\bx,\bx'\), the paired inputs have a fixed coupling satisfying \(\Var[Y(\bx,\xi)-Y(\bx',\xi')]\le\sigma^2\norm{\bx-\bx'}^{2\kappa}\), for constants \(\sigma\ge0\) and \(\kappa\in[0,1]\).
\end{assumption}

The coupling order \(\kappa\) is a property of the paired simulation experiment and determines the variance available to both classical and quantum mean estimation. Independent outputs with \(\sup_{\bx}\Var[Y(\bx,\xi)]\le\sigma^2/2\) satisfy Assumption~\ref{ass:pair} with \(\kappa=0\). Centered outputs that are Lipschitz in mean square satisfy \(\E[|(Y(\bx,\xi)-f(\bx))-(Y(\bx',\xi')-f(\bx'))|^2]\le\sigma^2\norm{\bx-\bx'}^2\), corresponding to \(\kappa=1\).

Coupled differences can retain useful variance reduction even when differentiating individual simulation responses does not recover the gradient of their expectation. For example, on \(\{\bx:\norm{\bx}\le1\}\), we let \(p(\bx)\mathrel{:=}1/4+\norm{\bx}^2/8+x_1^3/48\) and \(Y(\bx,\xi)\mathrel{:=}\mathbf1\{\xi\le p(\bx)\}\), where \(\xi\sim\operatorname{Unif}[0,1]\). Then, \(f=p\in C^\infty\) and \(\nabla^2f(\bx)\succeq\bI/8\), where \(\bI\) denotes the identity matrix. 
With a shared input, we have  \(\Var[Y(\bx,\xi)-Y(\bx',\xi)]\le|p(\bx)-p(\bx')|\le5\norm{\bx-\bx'}/16\), so Assumption~\ref{ass:pair} holds with \(\kappa=1/2\). 
Even though \(\nabla_{\bx}Y(\bx,\xi)=\bzero\) a.s. for every fixed \(\bx\), we still have \(\nabla f(\bx)\ne\bzero\) for \(\bx\ne\bx_*\).
Thus, averaging path-wise derivatives cannot recover the objective gradient in this example, while the paired differences still benefit from coupling.

Under Assumption~\ref{ass:pair}, we can use \(s_k=2^{\kappa-1}\sigma c_k^{\kappa-1}\) in Algorithm~\ref{alg:qace}. For feasible reconstruction~\eqref{eq:feasibleupdate}, we let \(R\ge\sup_{\bx,\bx'\in\mathcal X}\norm{\bx-\bx'}\) and use \(s^F=\sigma R^\kappa\sqrt{1+2^{4-2\kappa}}\,c^{\kappa-1}\). 
For a fixed initial decision \(\bx_0\), let \(\cF_k\) contain the information generated by the first \(k\) completed updates, with
\begin{equation*}
 \cF_k\mathrel{:=}
 \begin{cases}
  \sigma\{\bx_0,\boldsymbol\Delta_j,\widehat\mu_j:0\le j<k\},&\text{centered perturbations},\\
  \sigma\{\bx_0,\bx_j^{[i]},\widehat\mu_{j,i}^F:0\le j<k,\ 1\le i\le m\},&\text{feasible reconstruction}.
 \end{cases}
\end{equation*}
Both quantum implementations satisfy \(\E[(\widehat\mu_k-\mu_{c_k}(\bx_k,\boldsymbol\Delta_k))^2\mid\cF_k,\boldsymbol\Delta_k]\le\tau_k^2\). For feasible reconstruction~\eqref{eq:feasibleupdate}, we use conditionally independent scalar estimates given \(\cF_k,\bx_k^{[1]},\ldots,\bx_k^{[m]}\).

\subsection{Constant Step Sizes}
\label{sec:constant}
Let \(N_k^Q\) and \(N_k^C\) count all quantum and classical simulator queries used in \(k\) completed updates, respectively. 
We use geometrically decreasing perturbation sizes and estimation tolerances so that a constant step size continues to reduce the objective gap. Let \(\varepsilon_0>0\) be a constant, and let \(\mathcal C_2,\mathcal C_3\) denote the constants from Lemma~\ref{lem:propagation} in Appendix~\ref{app:proofs}. Throughout Section~\ref{sec:constant}, for centered perturbations~\eqref{eq:qaceupdate}, we take \(\eta=1/(L_1d)\), choose \(\varepsilon_0\) sufficiently large such that \(\varepsilon_0\ge(L_1/\rho)\E[f(\bx_0)-f_*]\), and set \(c_k=\min\{1,[\rho\varepsilon_0(1-\rho\eta/2)^k/(8\mathcal C_2L_2^2)]^{1/4}\}\) and \(\tau_k=\sqrt{\rho\varepsilon_0(1-\rho\eta/2)^k/(8\mathcal C_3d)}\). For feasible reconstruction~\eqref{eq:feasibleupdate}, take \(\eta=1/(2L_1)\), \(\varepsilon_0\ge\E[f(\bx_0)-f_*]\), \(c_k=\min\{1,[\rho\lambda\varepsilon_0(1-\rho\eta/2)^k/(8\mathcal C_2mL_2^2R^6)]^{1/4}\}\), and \(\tau_k=\sqrt{\rho\varepsilon_0(1-\rho\eta/2)^k/(8\mathcal C_3)}\).

\begin{theorem}
\label{thm:constant}
Under Assumptions~\ref{ass:objective} and \ref{ass:pair}, either procedure~\eqref{eq:qaceupdate} or~\eqref{eq:feasibleupdate} satisfies \(\lim_{k\to\infty}\bx_k=\bx_*\) a.s. 
For every \(0<\varepsilon<\varepsilon_0\), there exists an integer \(k_\varepsilon\) such that \(\E[f(\bx_{k_\varepsilon})-f_*]\le\varepsilon\). 
Centered perturbations~\eqref{eq:qaceupdate} attain
\begin{equation}
\label{eq:constantq}
 \E[N_{k_\varepsilon}^Q]=\cO(d^{3/2}\varepsilon^{-(3-\kappa)/4})
 \quad\text{and}\quad
 N_{k_\varepsilon}^C=\cO(d^2\varepsilon^{-(3-\kappa)/2})\quad\text{a.s.}
\end{equation}
Feasible reconstruction~\eqref{eq:feasibleupdate} retains these accuracy exponents. On the probability simplex \(\mathcal S_d\), vertex anchors yield
\begin{equation}
\label{eq:constantc}
 \E[N_{k_\varepsilon}^Q]=\cO(d^{(7-\kappa)/4}\varepsilon^{-(3-\kappa)/4})
 \quad\text{and}\quad
 N_{k_\varepsilon}^C=\cO(d^{(5-\kappa)/2}\varepsilon^{-(3-\kappa)/2})\quad\text{a.s.}
\end{equation}
The quantum bounds hold a.s. for the implementation with a deterministic query cap.
\end{theorem}

\begin{proof}{Proof of Theorem~\ref{thm:constant}.}
\label{app:constant}
Let \(\cE_k\mathrel{:=}\E[f(\bx_k)-f_*]\).
Lemma~\ref{lem:onestep} in Appendix~\ref{app:proofs} applies to \(L_1\norm{\bx_k-\bx_*}^2/2\) for centered perturbations~\eqref{eq:qaceupdate} and to \(f(\bx_k)-f_*\) for feasible reconstruction~\eqref{eq:feasibleupdate}. Both quantities start at most \(\varepsilon_0\), and the schedules bound the additive error in the expected recursion by \(\rho\eta\varepsilon_0(1-\rho\eta/2)^k/4\). Since
\begin{equation*}
 (1-\rho\eta)\varepsilon_0(1-\rho\eta/2)^k
 +\rho\eta\varepsilon_0(1-\rho\eta/2)^k/4
 \le\varepsilon_0(1-\rho\eta/2)^{k+1},
\end{equation*}
we obtain \(\cE_k\le\varepsilon_0(1-\rho\eta/2)^k\) by induction and smoothness. We define \(k_\varepsilon\mathrel{:=}\lceil\log(2\varepsilon_0/\varepsilon)/\log[1/(1-\rho\eta/2)]\rceil\); then \(\cE_{k_\varepsilon}\le\varepsilon\). Moreover, \(\sum_{k=0}^{\infty}\eta\varepsilon_0(1-\rho\eta/2)^k=2\varepsilon_0/\rho<\infty\), so Lemma~\ref{lem:propagation} implies \(\lim_{k\to\infty}\bx_k=\bx_*\) a.s.

For \(p\in\{1,2\}\), we use \(c_{j+1}\le c_j\) and \(\tau_{j+1}/\tau_j=\sqrt{1-\rho\eta/2}\) to obtain
\begin{equation*}
 \sum\nolimits_{j=0}^{k-1}\frac{1}{(c_j^{1-\kappa}\tau_j)^p}
 \le\frac{1}{[1-(1-\rho\eta/2)^{p/2}](c_{k-1}^{1-\kappa}\tau_{k-1})^p}
 \le\frac{4}{\rho\eta(c_{k-1}^{1-\kappa}\tau_{k-1})^p}.
\end{equation*}
At \(k=k_\varepsilon\), we have \(\varepsilon_0(1-\rho\eta/2)^{k-1}>\varepsilon/2\) and \(k_\varepsilon=\cO(\eta^{-1}\log(2\varepsilon_0/\varepsilon))\). For centered perturbations~\eqref{eq:qaceupdate}, \(\eta^{-1}=L_1d\), \(\tau_{k-1}^{-1}=\cO(\sqrt{d/\varepsilon})\), and \(c_{k-1}^{-(1-\kappa)}=\cO(\varepsilon^{-(1-\kappa)/4})\). Substitution into \eqref{eq:queryq} in Appendix~\ref{app:proofs} proves \eqref{eq:constantq}.

For feasible reconstruction~\eqref{eq:feasibleupdate}, we have \(\eta^{-1}=2L_1\), \(\tau_{k-1}^{-1}=\cO(1/\sqrt{\varepsilon})\), and \(c_{k-1}^{-(1-\kappa)}=\cO(\max\{1,mR^6/(\lambda\varepsilon)\}^{(1-\kappa)/4})\). Proposition~\ref{prop:querycost} therefore establishes the same accuracy exponents. On \(\mathcal S_d\), we use vertex anchors with \((m,\dim\mathcal T,R,\lambda)=(d,d-1,\sqrt2,1)\), which yields \eqref{eq:constantc}. Replacing \(\dim\mathcal T\) by \(m\) for the capped estimator leaves these orders unchanged.\Halmos
\end{proof}

The improvement comes from the cost of increasing precision. For fixed dimension, the finest stages use \(c\) of order \(\varepsilon^{1/4}\) and \(\tau\) of order \(\varepsilon^{1/2}\). Under the same coupling, both estimators use a standard-deviation bound of order \(c^{\kappa-1}\). Quantum estimation therefore costs \(\cO(\varepsilon^{-(3-\kappa)/4})\), while classical averaging squares this dependence. Summing the geometrically increasing costs over all preceding updates retains these exponents. Thus, coupling improves both rates, and increasing precision preserves the additional quadratic quantum improvement for every \(\kappa\in[0,1]\).

The same bounds specify when to stop under a fixed query budget \(N\). We choose a sufficiently large \(\mathcal C_4>0\). With the capped quantum estimator, the choice \(k=k_{\mathcal C_4N^{-4/(3-\kappa)}}\) ensures \(N_k^Q\le N\) a.s. and \(\E[f(\bx_k)-f_*]=\cO(N^{-4/(3-\kappa)})\) for all sufficiently large \(N\). For classical paired averaging, the choice \(k=k_{\mathcal C_4N^{-2/(3-\kappa)}}\) ensures \(N_k^C\le N\) a.s. and \(\E[f(\bx_k)-f_*]=\cO(N^{-2/(3-\kappa)})\).

\subsection{Diminishing Step Sizes}
\label{sec:diminishing}
Alternatively, we can set decreasing step sizes to control scalar estimation error. Throughout Section~\ref{sec:diminishing}, let \(\eta_k=\bar\eta(k+k_0)^{-\alpha}\) and \(c_k=\bar c(k+k_0)^{-\alpha/4}\), where \(\alpha\in(1/2,1)\), \(\bar\eta>0\), and \(0<\bar c\le1\). 
This polynomial form of step sizes is standard in the SA literature \citep[cf., e.g.,][]{hufu2025, huquantile2025, caocovar2025}. 
Choose an integer \(k_0\ge1\) sufficiently large such that \(\alpha k_0^{\alpha-1}\le\rho\bar\eta/2\) and \(\bar\eta k_0^{-\alpha}\le1/(L_1d)\) for centered perturbations~\eqref{eq:qaceupdate} or \(\bar\eta k_0^{-\alpha}\le1/L_1\) for feasible reconstruction~\eqref{eq:feasibleupdate}. 
For centered perturbations~\eqref{eq:qaceupdate}, we use the unbiased implementation of Lemma~\ref{lem:qvr} and set \(\tau_k=\bar\tau>0\).
For feasible reconstruction~\eqref{eq:feasibleupdate}, we set \(\tau_k=\bar\tau(k+k_0)^{-\alpha/2}\).

\begin{theorem}
\label{thm:diminishing}
Under Assumptions~\ref{ass:objective} and \ref{ass:pair}, either procedure~\eqref{eq:qaceupdate} or~\eqref{eq:feasibleupdate} satisfies \(\E[f(\bx_k)-f_*]=\cO((k+k_0)^{-\alpha})\), \(\E[\norm{\bx_k-\bx_*}^2]=\cO((k+k_0)^{-\alpha})\), and \(\lim_{k\to\infty}\bx_k=\bx_*\) a.s. For centered perturbations~\eqref{eq:qaceupdate}, we have
\begin{equation*}
 \E[N_k^Q]=\cO(k^{1+\alpha(1-\kappa)/4})
 \quad\text{and}\quad
 N_k^C=\cO(k^{1+\alpha(1-\kappa)/2})\quad\text{a.s.}
\end{equation*}
For feasible reconstruction~\eqref{eq:feasibleupdate}, the corresponding orders are \(\cO(k^{1+\alpha(3-\kappa)/4})\) and \(\cO(k^{1+\alpha(3-\kappa)/2})\).
\end{theorem}
For comparison, standard SPSA with one pair under independent noise uses scalar error tolerance \(\tau_k=\sigma/(2c_k)\). Balancing bias and noise through \(c_k=\bar c(k+k_0)^{-\alpha/6}\) yields \(\sqrt{\E[\norm{\bx_k-\bx_*}^2]}=\cO((k+k_0)^{-\alpha/3})\) \citep{hufu2025}, whereas Theorem~\ref{thm:diminishing} establishes \(\sqrt{\E[\norm{\bx_k-\bx_*}^2]}=\cO((k+k_0)^{-\alpha/2})\) for QACE. To account for the estimation effort behind this improvement, we choose a sufficiently small constant \(\mathcal C_5>0\). For centered perturbations~\eqref{eq:qaceupdate} with quantum estimation, we take \(k=\lfloor \mathcal C_5N^{4/[4+\alpha(1-\kappa)]}\rfloor\). For all sufficiently large \(N\), we have \(\E[N_k^Q]\le N\) and \(\E[f(\bx_k)-f_*]=\cO(N^{-4\alpha/[4+\alpha(1-\kappa)]})\). For classical averaging, we take \(k=\lfloor \mathcal C_5N^{2/[2+\alpha(1-\kappa)]}\rfloor\); then, we have \(N_k^C\le N\) a.s. and \(\E[f(\bx_k)-f_*]=\cO(N^{-2\alpha/[2+\alpha(1-\kappa)]})\).
\section{Query Lower Bounds}
\label{sec:lower}
We now establish lower bounds that match the accuracy exponents in Theorem~\ref{thm:constant} for fixed \(d\). Existing quantum lower bounds use multivariate mean recovery for stochastic gradient oracles \citep{sidford2023} and adversary arguments for best-arm identification \citep{wang2021quantumbandits}. We instead construct coupled simulators whose minimizers encode unknown coordinate signs. Their spatial scale links optimization error to the information in a paired query, allowing classical testing and quantum arguments based on state overlaps under the same coupling condition.

For an instance \(P\), we let \(f_P\) and \(\bx_{P,*}\) denote its objective and minimizer, respectively, and write \(f_{P,*}\mathrel{:=}f_P(\bx_{P,*})\). Let \(\mathcal N^C\) and \(\mathcal N^Q\) count all queries used by an arbitrary classical or quantum algorithm before it returns its decision.

\begin{theorem}
\label{thm:lower}
For every integer \(d\ge1\) and \(\kappa\in[0,1]\), there is a fixed class \(\mathcal P_{d,\kappa}\) satisfying Assumptions~\ref{ass:objective} and \ref{ass:pair} with \(\rho=1\), \(L_1=3\), \(L_2=1\), \(\sigma=2\), and \(\sup_{P\in\mathcal P_{d,\kappa}}[f_P(\bzero)-f_{P,*}]\le1\). 
Any algorithm returning \(\widehat\bx\) with \(\sup_{P\in\mathcal P_{d,\kappa}}\E_P[f_P(\widehat\bx)-f_{P,*}]\le\varepsilon\) obeys
\begin{equation*}
\sup\nolimits_{P\in\mathcal P_{d,\kappa}}\E_P[\mathcal N^C] =\Omega(d^{(1+\kappa)/2}\varepsilon^{-(3-\kappa)/2}) \quad\text{and}\quad \sup\nolimits_{P\in\mathcal P_{d,\kappa}}\E_P[\mathcal N^Q] =\Omega(d^{(3+\kappa)/4}\varepsilon^{-(3-\kappa)/4}),
\end{equation*}
for \(0<\varepsilon\le \mathcal C_6d^{-(1-\kappa)/(3-\kappa)}\), where \(\mathcal C_6>0\) is a constant. When \(\kappa=0\), the class \(\mathcal P_{d,0}\) can be chosen to use independent inputs within each output pair. There is also a class \(\mathcal P^\partial_\kappa\) on \(\mathcal S_3\) satisfying Assumptions~\ref{ass:objective} and \ref{ass:pair} with \((\rho,L_1,L_2,\sigma)=(1,3,1,2)\) and \((\bx_{P,*})_3=0\). For \(0<\varepsilon\le \mathcal C_6\), the criterion \(\sup_{P\in\mathcal P^\partial_\kappa}\E_P[f_P(\widehat\bx)-f_{P,*}]\le\varepsilon\) implies
\begin{equation*}
\sup\nolimits_{P\in\mathcal P^\partial_\kappa}\E_P[\mathcal N^C]=\Omega(\varepsilon^{-(3-\kappa)/2}) \quad\text{and}\quad \sup\nolimits_{P\in\mathcal P^\partial_\kappa}\E_P[\mathcal N^Q]=\Omega(\varepsilon^{-(3-\kappa)/4}).
\end{equation*}
At \(\kappa=0\), a fixed class \(\mathcal P_d^G\) with \((\rho,L_1,L_2,\sigma)=(1,3,1,2)\) and independent additive \(\operatorname{N}(0,1)\) noise satisfies
\begin{equation*}
 \sup\nolimits_{P\in\mathcal P_d^G}\E_P[\mathcal N^C]=\Omega(d^{3/2}\varepsilon^{-3/2}) \quad\text{for}\quad 0<\varepsilon\le \mathcal C_6d^{-1/3},
\end{equation*}
whenever \(\sup_{P\in\mathcal P_d^G}\E_P[f_P(\widehat\bx)-f_{P,*}]\le\varepsilon\).
\end{theorem}

\begin{proof}{Proof of Theorem~\ref{thm:lower}.}
\label{app:lower}
For each \(d\) and \(\kappa\), let \(\mathcal P_{d,\kappa}\) contain all instances indexed by a known scale \(0<\ell\le\min\{1/2,d^{-1/4},(4d^{1-\kappa/2})^{-1/(3-\kappa)}\}\) and an unknown \(\boldsymbol\nu\in\{-1,1\}^d\). Define \(\gamma_\ell\mathrel{:=}d^{1-\kappa/2}\ell^{3-\kappa}\) and draw \(\xi=(I,Z)\) according to \(\Prob_{\boldsymbol\nu}(I=i,Z=z)=(1+z\nu_i\gamma_\ell)/(2d)\), \(i\in\{1,\ldots,d\}\), \(z\in\{-1,1\}\). Sharing \(\xi\) within each pair, we set \(Y(\bx,\xi)\mathrel{:=}\norm{\bx}^2-\ell^\kappa d^{\kappa/2}\sin(x_I/\ell)Z\). Thus, \(f_{\ell,\boldsymbol\nu}(\bx)=\norm{\bx}^2-\ell^3\sum\nolimits_{i=1}^d\nu_i\sin(x_i/\ell)\).

Since \(\nabla^2 f_{\ell,\boldsymbol\nu}(\bx)=\operatorname{diag}(2+\nu_i\ell\sin(x_i/\ell))_{i=1}^d\), we have \(3\bI/2\preceq\nabla^2 f_{\ell,\boldsymbol\nu}(\bx)\preceq5\bI/2\) and \(\norm{\nabla^2 f_{\ell,\boldsymbol\nu}(\bx)-\nabla^2 f_{\ell,\boldsymbol\nu}(\bx')}\le\max_i|x_i-x_i'|\le\norm{\bx-\bx'}\). Hence, \(\rho=1\), \(L_1=3\), and \(L_2=1\) apply uniformly. Using \(\min\{4,t^2\}\le4^{1-\kappa}t^{2\kappa}\), \(t>0\), we obtain
\begin{equation*}
\begin{aligned}
 \Var[Y(\bx,\xi)-Y(\bx',\xi)]
 &\le\ell^{2\kappa}d^{\kappa-1}\sum\nolimits_i\min\{4,(x_i-x_i')^2/\ell^2\}\\
 &\le\ell^{2\kappa}d^\kappa\min\{4,\norm{\bx-\bx'}^2/(d\ell^2)\}
 \le4\norm{\bx-\bx'}^{2\kappa}.
\end{aligned}
\end{equation*}
Also, \(|\operatorname{supp}Y(\bx,\xi)|\le2d\) for every \(\bx\).

We let \(x_{\ell,*}\) solve \(2x_{\ell,*}=\ell^2\cos(x_{\ell,*}/\ell)\). For the one-dimensional objective, \(f'_{\ell,1}(\ell^2/4)=\ell^2[1/2-\cos(\ell/4)]<0\) and \(f'_{\ell,1}(\ell^2/2)=\ell^2[1-\cos(\ell/2)]\ge0\). Strict convexity and separability therefore imply \(\ell^2/4\le x_{\ell,*}\le\ell^2/2\) and \(\bx_*=x_{\ell,*}\boldsymbol\nu\). Moreover, \(\ell^3|\sin(t/\ell)|-t^2\le\ell^2|t|-t^2\le\ell^4/4\), so \(f(\bzero)-f_*\le d\ell^4/4\le1\).

To relate optimization error to sign recovery, we define \(\widehat\nu_i\mathrel{:=}2\mathbf1\{\widehat x_i\ge0\}-1\) and \(\mathcal H(\widehat{\boldsymbol\nu},\boldsymbol\nu)\mathrel{:=}\sum\nolimits_i\mathbf1\{\widehat\nu_i\ne\nu_i\}\). Then, \(\norm{\widehat\bx-\bx_*}^2\ge x_{\ell,*}^2\mathcal H(\widehat{\boldsymbol\nu},\boldsymbol\nu)\ge\ell^4\mathcal H(\widehat{\boldsymbol\nu},\boldsymbol\nu)/16\). By strong convexity, we can choose a constant \(C_2>0\) such that, with \(\ell=(C_3\varepsilon/d)^{1/4}\) and \(C_3\ge8/C_2\),
\begin{equation}
\label{eq:lowerhamming}
 f(\widehat\bx)-f_*\ge C_2\ell^4\mathcal H(\widehat{\boldsymbol\nu},\boldsymbol\nu) \quad\text{and}\quad \E_{\boldsymbol\nu}[\mathcal H(\widehat{\boldsymbol\nu},\boldsymbol\nu)]\le d/8.
\end{equation}
For sufficiently small \(\mathcal C_6\), we have \((C_3\varepsilon/d)^{1/4}\le\min\{1/2,d^{-1/4},(4d^{1-\kappa/2})^{-1/(3-\kappa)}\}\) and \(\gamma_\ell=(C_3\varepsilon)^{(3-\kappa)/4}d^{(1-\kappa)/4}\le1/4\).

\emph{Classical queries.} We grant the algorithm \(I,Z\) at each draw, from which it can compute the paired outputs at any decisions. Define \(\boldsymbol\nu^{(i)}\mathrel{:=}\boldsymbol\nu-2\nu_i\bunit_i\) and \(p_{\boldsymbol\nu,i}\mathrel{:=}\Prob_{\boldsymbol\nu}(\widehat\nu_i\ne\nu_i)\). Writing \(\mathcal L_{\boldsymbol\nu}\) for the law under \(\boldsymbol\nu\), the one-draw relative entropy is
\(\operatorname{KL}(\mathcal L_{\boldsymbol\nu}(I,Z)\Vert\mathcal L_{\boldsymbol\nu^{(i)}}(I,Z))=\gamma_\ell\log[(1+\gamma_\ell)/(1-\gamma_\ell)]/d\le4\gamma_\ell^2/d\).
For at most \(n\) adaptive draws, the chain rule and Pinsker's inequality imply \(p_{\boldsymbol\nu,i}+p_{\boldsymbol\nu^{(i)},i}\ge1-\sqrt{2n\gamma_\ell^2/d}\). Averaging this inequality and using \eqref{eq:lowerhamming}, we obtain \(n=\Omega(d/\gamma_\ell^2)\).

\emph{Quantum queries.} For \(\gamma\in[-1/4,1/4]\), define the planar rotation and the unknown oracle by
\begin{equation*}
 \mathbf R_\gamma\mathrel{:=} \begin{pmatrix} \sqrt{(1+\gamma)/2}&-\sqrt{(1-\gamma)/2}\\ \sqrt{(1-\gamma)/2}&\sqrt{(1+\gamma)/2} \end{pmatrix} \quad\text{and}\quad \mathcal U_{\boldsymbol\nu}\mathrel{:=}\sum\nolimits_{i=1}^d|i\rangle\langle i|\otimes\mathbf R_{\nu_i\gamma_\ell}.
\end{equation*}
Here \(|i\rangle\) stores index \(i\). Known operations prepare the uniform index state \((1/\sqrt d)\sum\nolimits_i|i\rangle\) and compute the paired outputs after applying \(\mathcal U_{\boldsymbol\nu}\). Thus, each simulator query, including inverse and controlled queries, uses one corresponding \(\mathcal U_{\boldsymbol\nu}\) query. To include forward, inverse, and inactive branches, we grant access to \(\mathcal V_{\boldsymbol\nu}\mathrel{:=}\operatorname{diag}(\bI,\mathcal U_{\boldsymbol\nu},\mathcal U_{\boldsymbol\nu}^\dagger)\).

By retaining measurements and randomization in auxiliary registers, we represent the algorithm after \(t\) queries by a unit vector \(|\Psi_{\boldsymbol\nu}^{(t)}\rangle\), including the known operations before the next query. Define \(\mathcal W_t\mathrel{:=}\sum\nolimits_{\boldsymbol\nu,i}\operatorname{Re}\langle\Psi_{\boldsymbol\nu}^{(t)}|\Psi_{\boldsymbol\nu^{(i)}}^{(t)}\rangle\). The initial state is independent of \(\boldsymbol\nu\), so \(\mathcal W_0=d2^d\). For the index projection \(\bP_i^{\rm idx}\mathrel{:=}\bI_{\rm control}\otimes|i\rangle\langle i|\otimes\bI_{\rm work}\),
\begin{equation*}
 (\mathcal V_{\boldsymbol\nu}^\dagger\mathcal V_{\boldsymbol\nu^{(i)}}-\bI)(\bI-\bP_i^{\rm idx})=0 \quad\text{and}\quad \norm{\mathcal V_{\boldsymbol\nu}^\dagger\mathcal V_{\boldsymbol\nu^{(i)}}-\bI} \le\arcsin\gamma_\ell\le2\gamma_\ell.
\end{equation*}
Since \(\sum\nolimits_{\boldsymbol\nu,i}\norm{\bP_i^{\rm idx}\Psi_{\boldsymbol\nu}^{(t)}}^2=2^d\), the Cauchy--Schwarz inequality yields
\begin{equation*}
 |\mathcal W_{t+1}-\mathcal W_t|
 \le2\gamma_\ell\sum\nolimits_{\boldsymbol\nu,i}\norm{\bP_i^{\rm idx}\Psi_{\boldsymbol\nu}^{(t)}}\norm{\bP_i^{\rm idx}\Psi_{\boldsymbol\nu^{(i)}}^{(t)}}
 \le2\gamma_\ell2^d.
\end{equation*}
Known operations preserve the overlaps because \(\langle V\Psi|V\Psi'\rangle=\langle\Psi|\Psi'\rangle\) for unitary \(V\).

At termination after at most \(n\) queries, projection onto the subspaces for the decoded bit and the Cauchy--Schwarz inequality imply \(|\langle\Psi_{\boldsymbol\nu}^{(n)}|\Psi_{\boldsymbol\nu^{(i)}}^{(n)}\rangle|\le\sqrt{p_{\boldsymbol\nu,i}}+\sqrt{p_{\boldsymbol\nu^{(i)},i}}\). Concavity of the square root and \eqref{eq:lowerhamming} imply \(\mathcal W_n\le2d2^d\sqrt{(d2^d)^{-1}\sum\nolimits_{\boldsymbol\nu,i}p_{\boldsymbol\nu,i}}\le d2^d/\sqrt2\). Consequently, \((1-1/\sqrt2)d2^d\le\mathcal W_0-\mathcal W_n\le2n\gamma_\ell2^d\), and \(n=\Omega(d/\gamma_\ell)\).

For a random query count \(\mathcal N\) with \(\sup_{\boldsymbol\nu}\E_{\boldsymbol\nu}[\mathcal N]\le n\), truncate after \(\lceil C_4n\rceil\) queries, where \(C_4>8\), and return an arbitrary sign vector upon truncation. Markov's inequality implies \(\sup_{\boldsymbol\nu}\Prob_{\boldsymbol\nu}(\mathcal N>\lceil C_4n\rceil)\le1/C_4\), so the truncated output \(\widetilde{\boldsymbol\nu}\) satisfies \(\sup_{\boldsymbol\nu}\E_{\boldsymbol\nu}[\mathcal H(\widetilde{\boldsymbol\nu},\boldsymbol\nu)]\le d(1/8+1/C_4)<d/4\). The testing and overlap arguments still apply, since \(1-2(1/8+1/C_4)>0\) and \(2\sqrt{1/8+1/C_4}<1\). Under a uniform prior on \(\boldsymbol\nu\), \(\E[\mathcal H(\widehat{\boldsymbol\nu},\boldsymbol\nu)\mathbf1\{\mathcal N=0\}]=(d/2)\Prob(\mathcal N=0)\). Hence, from \eqref{eq:lowerhamming}, we obtain \(\Prob(\mathcal N\ge1)\ge3/4\), \(n\ge3/4\), and \(\lceil C_4n\rceil\le(C_4+4/3)n\). We obtain \(n=\Omega(d/\gamma_\ell^2)\) classically and \(n=\Omega(d/\gamma_\ell)\) quantumly. Substituting the selected \(\gamma_\ell\) proves the first two bounds.

At \(\kappa=0\), replacing the shared input by independent copies \(\xi,\xi'\) preserves \(f_{\ell,\boldsymbol\nu}\) and ensures \(\Var[Y(\bx,\xi)-Y(\bx',\xi')]\le2\); each paired query then uses two draws classically and two corresponding queries to \(\mathcal U_{\boldsymbol\nu}\), its inverse, or its controlled version quantumly, so the preceding lower bounds and truncation argument change only by constant factors.

\emph{Independent Gaussian responses.} For \(\mathcal P_d^G\), retain the same objectives and use \(Y(\bx,\xi)=f_{\ell,\boldsymbol\nu}(\bx)+\xi\), with \(\{(\xi_t,\xi_t')\}_{t\ge1}\overset{\mathrm{iid}}{\sim}\operatorname{N}(\bzero,\bI_2)\). The Gaussian inputs remain internal to the simulator, and each classical query returns only the output pair. Then, \(\Var[Y(\bx,\xi)-Y(\bx',\xi')]=2\) and \(\sup_{\bx}|f_{\ell,\boldsymbol\nu}(\bx)-f_{\ell,\boldsymbol\nu^{(i)}}(\bx)|\le2\ell^3\). For the output pair \(\mathbf Y=(Y(\bx,\xi),Y(\bx',\xi'))\), we have \(\operatorname{KL}(\mathcal L_{\boldsymbol\nu}(\mathbf Y)\Vert\mathcal L_{\boldsymbol\nu^{(i)}}(\mathbf Y))\le4\ell^6\). Applying the chain rule and Pinsker's inequality, we obtain \(p_{\boldsymbol\nu,i}+p_{\boldsymbol\nu^{(i)},i}\ge1-\sqrt{2n\ell^6}\), and \eqref{eq:lowerhamming} implies \(n=\Omega(\ell^{-6})=\Omega(d^{3/2}\varepsilon^{-3/2})\). The truncation argument therefore yields \(\sup_{P\in\mathcal P_d^G}\E_P[\mathcal N^C]=\Omega(d^{3/2}\varepsilon^{-3/2})\).

\emph{Boundary simplex.} For \(\mathcal P^\partial_\kappa\) on \(\mathcal S_3\), define \(z(\bx)\mathrel{:=}(x_1-x_2)/\sqrt2\) and \(Y(\bx,\xi)\mathrel{:=}\norm{\bx}^2+2x_3-\ell^\kappa\sin(z(\bx)/\ell)\xi\), where \(\Prob_\nu(\xi=a)=(1+a\nu\ell^{3-\kappa})/2\), \(a,\nu\in\{-1,1\}\). Since \(|z(\bx)-z(\bx')|\le\norm{\bx-\bx'}\), we obtain \(\Var[Y(\bx,\xi)-Y(\bx',\xi)]\le4\norm{\bx-\bx'}^{2\kappa}\). Moreover, \(f(\bx)=1/2+f_{\ell,\nu}(z(\bx))+x_3+3x_3^2/2\), so \(\bx_*=(1/2+\nu x_{\ell,*}/\sqrt2,\ 1/2-\nu x_{\ell,*}/\sqrt2,\ 0)\) and \(f(\widehat\bx)-f_*\ge f_{\ell,\nu}(z(\widehat\bx))-f_{\ell,\nu}(\nu x_{\ell,*})\). Each simplex query uses one scalar pair at the corresponding \(z\) values and known arithmetic. Applying the \(d=1\) bounds, we obtain \(\sup_{P\in\mathcal P^\partial_\kappa}\E_P[\mathcal N^C]=\Omega(\varepsilon^{-(3-\kappa)/2})\) and \(\sup_{P\in\mathcal P^\partial_\kappa}\E_P[\mathcal N^Q]=\Omega(\varepsilon^{-(3-\kappa)/4})\).\Halmos
\end{proof}

Together, Theorems~\ref{thm:constant} and \ref{thm:lower} establish the minimax optimal exponents \((3-\kappa)/2\) and \((3-\kappa)/4\) in the classical and quantum query complexities, respectively, for fixed decision dimension. The classical lower bound permits adaptive evaluation points and batch sizes. Consequently, the quantum improvement cannot be recovered by reallocating classical simulation effort or changing the search rule within the stated oracle model. The dimension bounds differ; at \(\kappa=0\), the classical lower bound with independent noise is \(\Omega(d^{3/2}\varepsilon^{-3/2})\), compared with the upper bound \(\cO(d^2\varepsilon^{-3/2})\).
\section{Numerical Experiments}
\label{sec:numerical}
In this section, we examine the query cost of estimating a paired-difference mean and then study how this saving affects optimization. For centered perturbations on \(\R^2\), we compare QACE and its classical counterpart on an objective with quadratic and sinusoidal terms, observed through discontinuous indicator responses under independent and shared inputs. For feasible reconstruction, we consider allocating one unit of resource among three channels with exponential demands, penalties for shortage events, and a quadratic allocation cost. The optimal allocation assigns no resource to the third channel. 
The quantum implementation uses range-normalized iterative quantum amplitude estimation (IQAE) \citep{grinko2021iterative}, while its classical counterpart replaces mean estimation by averaging within the same decision updates. 
The electronic companion verifies that the IQAE estimator meets the prescribed scalar RMSE tolerance, as required in Lemma~\ref{lem:onestep}. The theoretical query bounds use the variance-sensitive estimator of Lemma~\ref{lem:qvr}.
Quantum measurement outcomes are simulated classically, and effort is measured in oracle queries. 

Figures~\ref{fig:numerical}(a)--\ref{fig:numerical}(b) show the queries required to attain a target mean optimality gap, while Figures~\ref{fig:numerical}(c)--\ref{fig:numerical}(d) report RMSE and mean optimality gap against oracle queries, respectively.
Experimental settings, confidence intervals, and query accounting appear in the electronic companion.
In the mean estimation experiment, the empirical RMSE decreases approximately as \(\bar Q^{-0.512}\) for classical averaging and \(\bar Q^{-0.998}\) for IQAE, where \(\bar Q\) denotes the average query count. These rates are close to inverse square root and inverse dependence on the query count, respectively. The optimization experiments show how this improvement carries through the decision updates. For centered perturbations~\eqref{eq:qaceupdate} on \(\R^2\) with independent inputs, the first tested budgets attaining the empirical target \(\varepsilon=10^{-6}\) are \(3.16\times10^7\) classical queries and \(5.62\times10^5\) quantum queries, a ratio of about \(56.2\). For resource allocation with a boundary optimum, the corresponding budgets at \(\varepsilon=10^{-7}\) are \(5.62\times10^9\) classical queries and \(5.62\times10^7\) quantum queries, achieving a \(100\)-fold improvement in query efficiency.

\newsavebox{\boxfigQACE}
\sbox{\boxfigQACE}{%
\begin{minipage}{0.8\textwidth}\centering
\begin{minipage}[t]{0.48\linewidth}\centering
\includegraphics[width=0.98\linewidth]{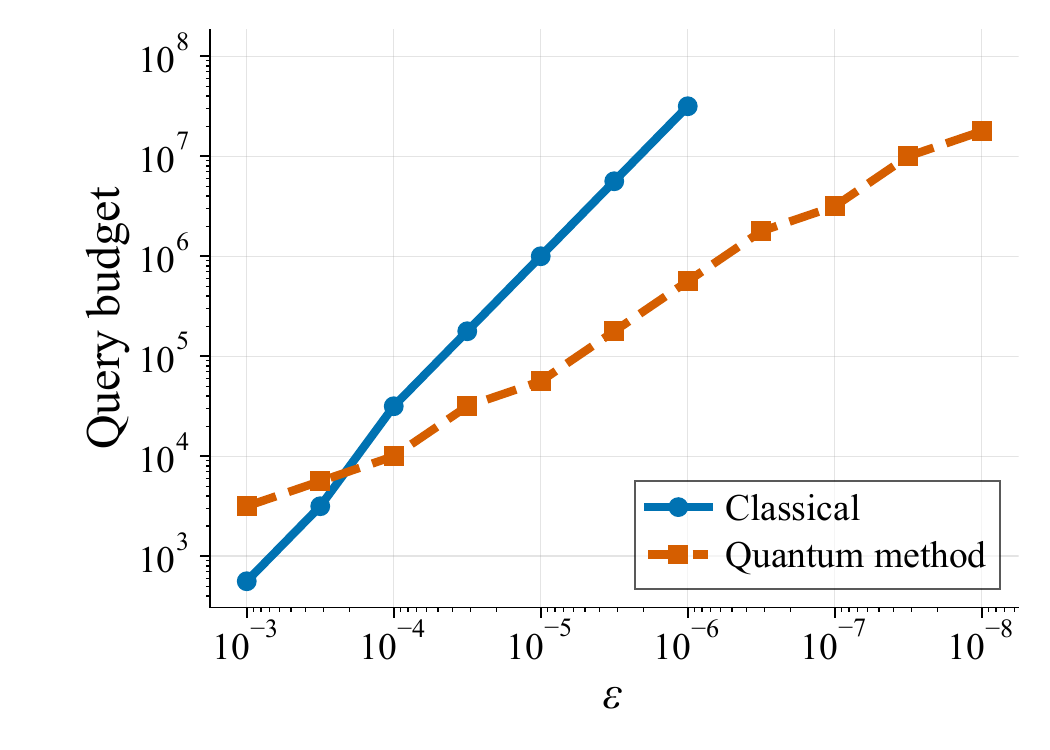}\par
\scriptsize (a) Centered perturbations, independent inputs.
\end{minipage}%
\hfill
\begin{minipage}[t]{0.48\linewidth}\centering
\includegraphics[width=0.98\linewidth]{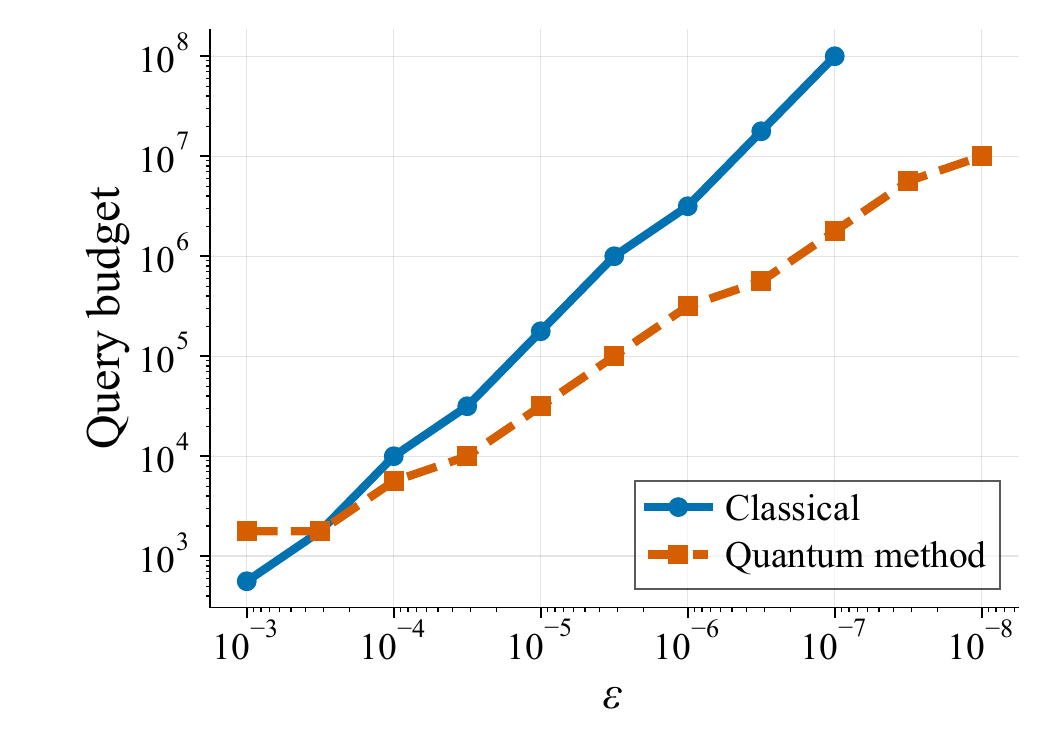}\par
\scriptsize (b) Centered perturbations, shared inputs.
\end{minipage}\par
\begin{minipage}[t]{0.48\linewidth}\centering
\includegraphics[width=0.98\linewidth]{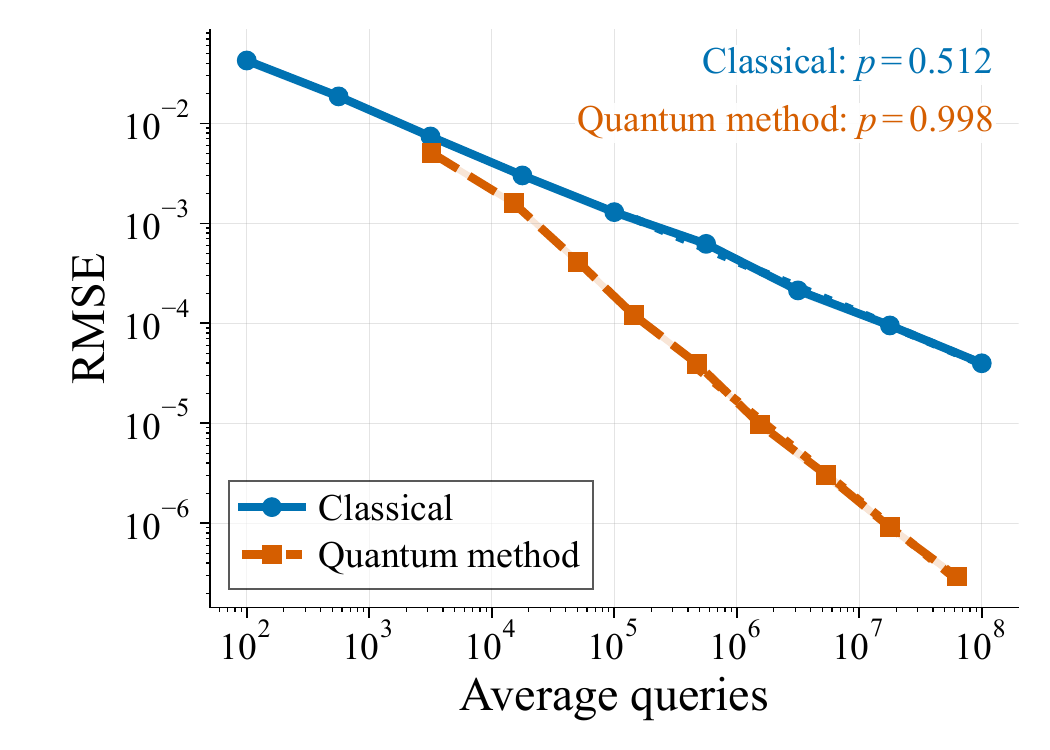}\par
\scriptsize (c) Paired-difference mean estimation.
\end{minipage}%
\hfill
\begin{minipage}[t]{0.48\linewidth}\centering
\includegraphics[width=0.98\linewidth]{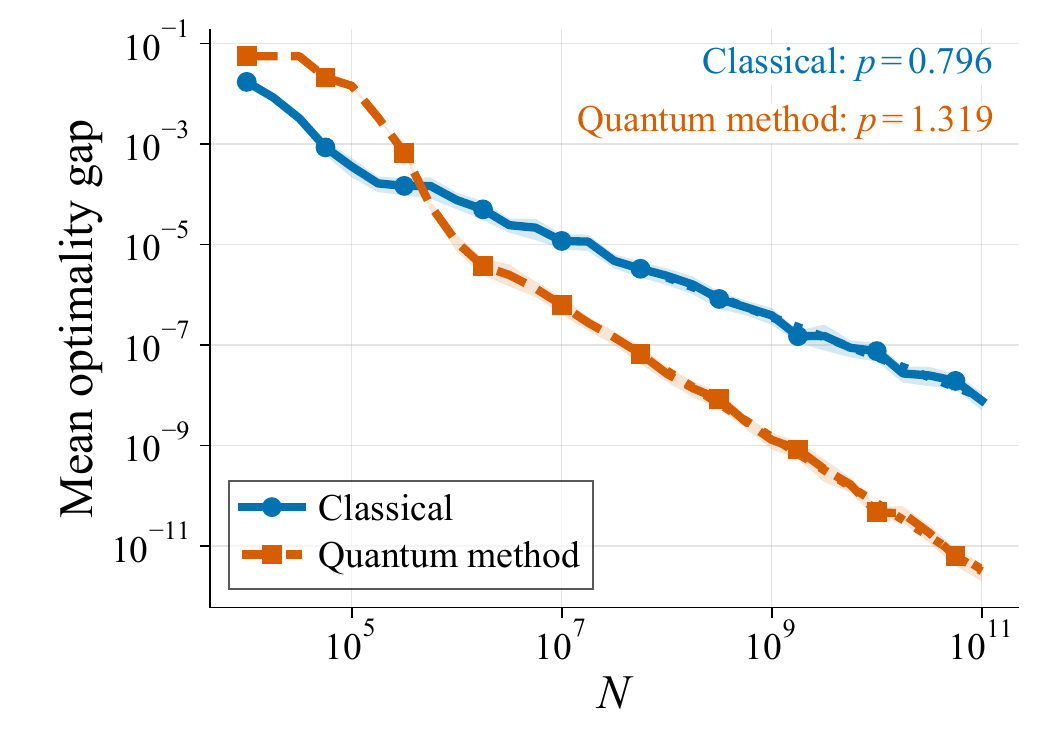}\par
\scriptsize (d) Resource allocation at a boundary optimum.
\end{minipage}
\end{minipage}}

\begin{figure}
\FIGURE
{\usebox{\boxfigQACE}}
{Simulation effort and accuracy with classical averaging and quantum mean estimation.\label{fig:numerical}}{}
\end{figure}

\section{Conclusions}
\label{sec:conclusion}
Quantum mean estimation reduces the query complexity of gradient-based search for smooth strongly convex expected objectives. By coordinating estimation accuracy with shrinking perturbations, QACE carries the saving in each gradient estimate through the full sequence of decision updates. Coupling improves both classical and quantum rates while preserving their quadratic separation, and matching lower bounds establish minimax optimal accuracy exponents for fixed decision dimension. These bounds show why adaptive classical batching cannot reproduce the quantum gain. Feasible reconstruction retains the exponents when every simulation query must be feasible, including at boundary optima.
\begin{APPENDICES}
\renewcommand{\theHsection}{appendix.\Alph{section}}
\section{Proofs of Auxiliary Results}
\label{app:proofs}
\begin{proof}{Proof of Lemma~\ref{lem:qvr}.}
For a finite encoding, write
\(
\mathcal A|0\rangle
=\sum\nolimits_z\sqrt{\Prob[Y=z]}\,|z\rangle|\mathrm{aux}_z\rangle,
\)
where \(|z\rangle\) stores the encoded output and \(|\mathrm{aux}_z\rangle\) is a normalized auxiliary state retained for reversibility.
For \(\Var(Y)\le s^2\) and \(a>0\), quantum mean estimation returns \(\widehat\mu\) satisfying 
\(
\Prob(|\widehat\mu-\E[Y]|\le a)\ge2/3
\)
and 
\(
\mathsf q(\widehat\mu)\le C_1(1+s/a)
\)
a.s., for a constant \(C_1>0\). 
Theorem~1.1 of \citet{kothari2023} applies to finitely encoded outputs with exact reversible value arithmetic; for continuous laws, we use an ideal real arithmetic model with this guarantee. 

If \(\tau\ge s\), take \(\widehat\mu=Y\); then \(\E[\widehat\mu]=\E[Y]\), \(\E[(\widehat\mu-\E[Y])^2]\le\tau^2\), and \(\mathsf q(\widehat\mu)=1\). We therefore consider \(0<\tau<s\). Fix an integer \(J\ge1\) and take \(\widehat\mu^{(0)}\) to be the median of \(4J+3\) independent copies of \(Y\). For \(t\ge4s\), Chebyshev's inequality and a union bound imply \(\Prob[|\widehat\mu^{(0)}-\E[Y]|>t]\le2^{4J+3}(s/t)^{4J+4}\). Integrating this tail bound, we obtain \(\E[(\widehat\mu^{(0)}-\E[Y])^2\mathbf1\{|\widehat\mu^{(0)}-\E[Y]|>4s\}]\le C_1s^22^{-4J}\) for a sufficiently large constant \(C_1>0\).

For \(j=1,\ldots,J\), let \(T_j\) be the median of independent quantum estimates chosen so that \(\Prob[|T_j-\E[Y]|>4s\,2^{-j}]\le2^{-4(J-j+1)}\) and \(\mathsf q(T_j)\le C_1(1+2^j)(J-j+1)\). Define \(\widehat\mu^{(j)}=\Pi_{[\widehat\mu^{(j-1)}-4s\,2^{1-j},\widehat\mu^{(j-1)}+4s\,2^{1-j}]}(T_j)\). If \(|\widehat\mu^{(j-1)}-\E[Y]|\le 4s\,2^{1-j}\) and \(|T_j-\E[Y]|\le 4s\,2^{-j}\), then \(|\widehat\mu^{(j)}-\E[Y]|\le 4s\,2^{-j}\). If \(|\widehat\mu^{(0)}-\E[Y]|\le4s\), the first \(j\) with \(|T_j-\E[Y]|>4s\,2^{-j}\) implies \(|\widehat\mu^{(J)}-\E[Y]|\le 4s\,2^{1-j}+\sum_{t=j}^J 4s\,2^{1-t}<12s\,2^{1-j}\); otherwise, \(|\widehat\mu^{(J)}-\E[Y]|\le|\widehat\mu^{(0)}-\E[Y]|+8s\le3|\widehat\mu^{(0)}-\E[Y]|\). Partitioning by the first failure, we obtain
\begin{equation*}
 \E[(\widehat\mu^{(J)}-\E[Y])^2]
 \le 16s^2 2^{-2J}+9\sum\nolimits_{j=1}^J (4s\,2^{1-j})^2\,2^{-4(J-j+1)}+C_1s^22^{-4J}
 \le C_1s^22^{-2J}.
\end{equation*}
Moreover, \(\mathsf q(\widehat\mu^{(J)})\le C_1[J+\sum_{j=1}^J(1+2^j)(J-j+1)]\le C_2 2^J\). Thus, we set \(J=\lceil\log_2(\sqrt{C_1}s/\zeta)\rceil\) and \(\widehat\mu_\zeta=\widehat\mu^{(J)}\) to obtain \(\E[(\widehat\mu_\zeta-\E[Y])^2]\le\zeta^2\) and \(\mathsf q(\widehat\mu_\zeta)\le C_2s/\zeta\) for \(0<\zeta<s\).

For a constant \(C_2\ge1\) to be chosen, draw \(J_*\ge1\) independently with \(\Prob[J_*=j]=2/3^j\). Using independent estimates, let \(W_j=\widehat\mu_{\tau2^{-j}/C_2}-\widehat\mu'_{\tau2^{1-j}/C_2}\) and return \(\widehat\mu=\widehat\mu_{\tau/C_2}+3^{J_*}W_{J_*}/2\). The moment bounds and \(\sum_{j\ge1}3^j4^{-j}/2=3/2\) imply \(\sum_{j\ge1}3^j\E[W_j^2]/2\le C_1\tau^2/C_2^2\) and \(\sum_{j\ge1}\E[|W_j|]<\infty\). Hence, \(\E[\widehat\mu]=\E[\widehat\mu_{\tau/C_2}]+\sum_{j\ge1}\E[W_j]=\E[Y]\). Moreover,
\(\E[(\widehat\mu-\E[Y])^2]\le2\tau^2/C_2^2+\sum_{j\ge1}3^j\E[W_j^2]\le2(1+C_1)\tau^2/C_2^2\le\tau^2\)
for \(C_2^2\ge2(1+C_1)\). Since \(\sum_{j\ge1}(2/3)^j=2\), we also have \(\E[\mathsf q(\widehat\mu)]\le C_1(s/\tau)[1+\sum_{j\ge1}(2/3)^j]=\cO(s/\tau)\).

For an encoded output \(\widetilde Y\) coupled to \(Y\) with \(\E[(\widetilde Y-Y)^2]\le\tau^2/4\), we have \(|\E[\widetilde Y]-\E[Y]|\le\tau/2\) and \(\operatorname{sd}(\widetilde Y)\le s+\tau/2\). Applying \(\widehat\mu_{\tau/2}\) to \(\widetilde Y\) with this standard-deviation bound, we obtain
\(\sqrt{\E[(\widehat\mu_{\tau/2}-\E[Y])^2]}\le\sqrt{\E[(\widehat\mu_{\tau/2}-\E[\widetilde Y])^2]}+|\E[\widetilde Y]-\E[Y]|\le\tau\)
and \(\mathsf q(\widehat\mu_{\tau/2})\le C_1(1+s/\tau)\).

For the event example following Assumption~\ref{ass:pair}, let \(b_1\ge1\) be the number of bits used to represent the input and threshold. Take \(\widetilde\xi=2^{-b_1}\lfloor2^{b_1}\xi\rfloor\) and \(|\widetilde p(\bx)-p(\bx)|\le2^{-b_1}\). Then, \(\Prob(\mathbf1\{\widetilde\xi\le\widetilde p(\bx)\}\ne Y(\bx,\xi))\le C_1 2^{-b_1}\). We round each scalar \(z\) to \(\widetilde z\) with \(|\widetilde z-z|\le\tau/4\). For the complete scalar output \(Y\) of the Richardson sampler and its encoded value \(\widetilde Y\), this yields \(\sqrt{\E[(\widetilde Y-Y)^2]}\le C_1 2^{-b_1/2}/c+\tau/4\le\tau/2\) whenever \(2^{b_1}\ge C_2/(c^2\tau^2)\). For fixed \(d\), the schedules in Section~\ref{sec:constant} imply \(b_1,b_2=\cO(\log(1/\varepsilon))\), where \(b_2\) is the number of bits used to store the rounded scalar.\Halmos
\end{proof}

\begin{lemma}
\label{lem:onestep}
Define 
\(
V_k:=L_1\norm{\bx_k-\bx_*}^2/2
\)
for the centered perturbations~\eqref{eq:qaceupdate}, and
\(
V_k:=f(\bx_k)-f_*
\)
for the feasible reconstruction~\eqref{eq:feasibleupdate}.
Under Assumptions~\ref{ass:objective} and \ref{ass:pair}, there exist constants \(\mathcal C_2,\mathcal C_3>0\), depending only on \(L_1/\rho\), such that the centered update \eqref{eq:qaceupdate}, with unbiased scalar estimates and \(\eta_k\le1/(L_1d)\), satisfies
\begin{equation}
\label{eq:qacerecursion}
 \E[V_{k+1}\mid\cF_k]
 \le(1-\rho\eta_k)V_k+\mathcal C_2L_2^2\eta_kc_k^4+\mathcal C_3L_1d^2\eta_k^2\tau_k^2.
\end{equation}
For possibly biased scalar estimates, we have 
\(
\E[V_{k+1}\mid\cF_k]
 \le(1-\rho\eta_k)V_k+\mathcal C_2L_2^2\eta_kc_k^4+\mathcal C_3d\eta_k\tau_k^2.
\)
The feasible update \eqref{eq:feasibleupdate}, with \(\eta_k\le1/L_1\), satisfies
\begin{equation}
\label{eq:feasiblerecursion}
 \E[V_{k+1}\mid\cF_k]
 \le(1-\rho\eta_k)V_k+\mathcal C_2\frac{mL_2^2R^6}{\lambda}\eta_kc_k^4+\mathcal C_3\eta_k\tau_k^2,
\end{equation}
using scalar tolerances \(\tau_k\sqrt{\lambda/\dim\mathcal T}\) for unbiased estimates and \(\tau_k\sqrt{\lambda/m}\) for possibly biased estimates.
\end{lemma}

\begin{proof}{Proof of Lemma~\ref{lem:onestep}.}
\label{app:onestep}
\emph{Centered perturbations~\eqref{eq:qaceupdate}.} Conditional on \(\cF_k\), write \(\bx=\bx_k\), \(c=c_k\), \(\eta=\eta_k\), and \(\tau=\tau_k\). For \(\boldsymbol\Delta\in[-1,1]^d\), define
\begin{equation*}
 \mathcal R_c(\boldsymbol\Delta)
 :=\frac{f(\bx+c\boldsymbol\Delta/\sqrt d)-f(\bx-c\boldsymbol\Delta/\sqrt d)}{2c}
 -\frac{\ip{\nabla f(\bx)}{\boldsymbol\Delta}}{\sqrt d}.
\end{equation*}
Taylor's formula and Assumption~\ref{ass:objective} imply
\( |\mathcal R_c(\boldsymbol\Delta)|\le L_2c^2\norm{\boldsymbol\Delta}^3/(6d^{3/2})\),
\(\norm{\nabla_{\boldsymbol\Delta}\mathcal R_c(\boldsymbol\Delta)}\le L_2c^2\norm{\boldsymbol\Delta}^2/(2d^{3/2})\), and
\(\norm{\nabla_{\boldsymbol\Delta}^2\mathcal R_c(\boldsymbol\Delta)}\le L_2c^2\norm{\boldsymbol\Delta}/d^{3/2}\).
For the sampled direction \(\boldsymbol\Delta=\boldsymbol\Delta_k\), \(\mathcal R_c(-\boldsymbol\Delta)=-\mathcal R_c(\boldsymbol\Delta)\) implies \(\E[\mathcal R_c(\boldsymbol\Delta)\mid\cF_k]=0\).

Let \(\boldsymbol\Delta^{(i)}\) denote \(\boldsymbol\Delta\) with sign \(i\) flipped. Since \(\norm{t\boldsymbol\Delta+(1-t)\boldsymbol\Delta^{(i)}}\le\sqrt d\) for \(t\in[0,1]\), Taylor's formula yields
\(|\mathcal R_c(\boldsymbol\Delta)-\mathcal R_c(\boldsymbol\Delta^{(i)})|\le2|\partial_i\mathcal R_c(\boldsymbol\Delta)|+2L_2c^2/d\).
Therefore, we have from the Efron--Stein inequality that
\begin{equation*}
\begin{aligned}
 \E[\mathcal R_c(\boldsymbol\Delta)^2\mid\cF_k]
 &\le\frac14\sum\nolimits_i\E[(\mathcal R_c(\boldsymbol\Delta)-\mathcal R_c(\boldsymbol\Delta^{(i)}))^2\mid\cF_k]\\
 &\le2\E[\norm{\nabla_{\boldsymbol\Delta}\mathcal R_c(\boldsymbol\Delta)}^2\mid\cF_k]+2L_2^2c^4/d
 \le\frac{5L_2^2c^4}{2d}.
\end{aligned}
\end{equation*}
Since \(\E[\boldsymbol\Delta\boldsymbol\Delta^\top\mid\cF_k]=\bI\),
\(\norm{\E[\mathcal R_c(\boldsymbol\Delta)\boldsymbol\Delta\mid\cF_k]}^2\le\E[\mathcal R_c(\boldsymbol\Delta)^2\mid\cF_k]\).
For unbiased scalar estimates, the conditional mean and second moment of \(\bg_k\) are 
\(
\E[\bg_k\mid\cF_k]=\nabla f(\bx)+\sqrt d\,\E[\mathcal R_c(\boldsymbol\Delta)\boldsymbol\Delta\mid\cF_k]
\)
and 
\(
\E[\norm{\bg_k}^2\mid\cF_k]
 \le d\norm{\nabla f(\bx)}^2
 +2d^{3/2}\ip{\nabla f(\bx)}{\E[\mathcal R_c(\boldsymbol\Delta)\boldsymbol\Delta\mid\cF_k]}
 +d^2(\E[\mathcal R_c(\boldsymbol\Delta)^2\mid\cF_k]+\tau^2).
\)

Because \(\nabla f(\bx_*)=\bzero\) and \(\eta d\le1/L_1\), the Hessian bounds imply
\begin{equation*}
 \norm{\bx-\bx_*-\eta d\nabla f(\bx)}
 =\Big\|\Big[\bI-\eta d\int_0^1\nabla^2f(\bx_*+t(\bx-\bx_*))\,dt\Big](\bx-\bx_*)\Big\|
 \le(1-\rho\eta d)\norm{\bx-\bx_*}.
\end{equation*}
By nonexpansiveness of projection and the two moment bounds, we obtain
\begin{equation*}
\begin{aligned}
 \E[\norm{\bx_{k+1}-\bx_*}^2\mid\cF_k]
 &\le(1-1/d)\norm{\bx-\bx_*}^2
 +\norm{\bx-\bx_*-\eta d\nabla f(\bx)}^2/d\\
 &\quad-2\eta\sqrt d\,\ip{\bx-\bx_*-\eta d\nabla f(\bx)}{\E[\mathcal R_c(\boldsymbol\Delta)\boldsymbol\Delta\mid\cF_k]}
 +\eta^2d^2\big(\E[\mathcal R_c(\boldsymbol\Delta)^2\mid\cF_k]+\tau^2\big).
\end{aligned}
\end{equation*}
The absolute value of the inner product term is at most
\begin{equation*}
\begin{aligned}
 2\eta(1-\rho\eta d)\norm{\bx-\bx_*}
 \sqrt{d\E[\mathcal R_c(\boldsymbol\Delta)^2\mid\cF_k]}
 \le\rho\eta(1-\rho\eta d)\norm{\bx-\bx_*}^2
 +\frac{\eta d(1-\rho\eta d)}{\rho}\E[\mathcal R_c(\boldsymbol\Delta)^2\mid\cF_k].
\end{aligned}
\end{equation*}
Here the squared-distance coefficient simplifies to
\(1-1/d+(1-\rho\eta d)^2/d+\rho\eta(1-\rho\eta d)=1-\rho\eta\), while
\(\eta d(1-\rho\eta d)/\rho+\eta^2d^2=\eta d/\rho\). Thus, we have
\begin{equation*}
\begin{aligned}
 \E[\norm{\bx_{k+1}-\bx_*}^2\mid\cF_k]
 &\le(1-\rho\eta)\norm{\bx-\bx_*}^2
 +\frac{\eta d}{\rho}\E[\mathcal R_c(\boldsymbol\Delta)^2\mid\cF_k]+\eta^2d^2\tau^2\\
 &\le(1-\rho\eta)\norm{\bx-\bx_*}^2+\frac{5L_2^2}{2\rho}\eta c^4+\eta^2d^2\tau^2.
\end{aligned}
\end{equation*}
Multiplying by \(L_1/2\) proves \eqref{eq:qacerecursion}, with \(\mathcal C_2=5L_1/(4\rho)\) and \(\mathcal C_3=1/2\) sufficient for this case.

For possibly biased scalar estimates, replace \(\mathcal R_c(\boldsymbol\Delta)\) in the preceding calculation by
\(\mathcal R_c(\boldsymbol\Delta)+\widehat\mu_k-\mu_c(\bx,\boldsymbol\Delta)\).
Its second moment is at most \(2\E[\mathcal R_c(\boldsymbol\Delta)^2\mid\cF_k]+2\tau^2\). Applying the same projection and Young bounds, we obtain
\(
\E[V_{k+1}\mid\cF_k]
 \le(1-\rho\eta)V_k+5L_1L_2^2\eta c^4/(2\rho)+L_1d\eta\tau^2/\rho.
\)
Since \(\norm{\nabla f(\bx)}\le L_1\norm{\bx-\bx_*}\), the conditional moment bounds and nonexpansiveness of projection give \(\E[\norm{\bx_k-\bx_*}^2+V_k+\norm{\bg_k}^2]<\infty\) by induction from the deterministic initialization \(\bx_0\).

\emph{Feasible reconstruction~\eqref{eq:feasibleupdate}.} Conditional on \(\cF_k,\bx_k^{[1]},\ldots,\bx_k^{[m]}\), we suppress the iteration indices and conditioning in the following expectations and variances. By spectral decomposition of \(\bM_{\bx}\), we have
\(\sum_i\bs_i\bw_i^\top=\bP\), \(\sum_i\norm{\bs_i}^2=\operatorname{tr}(\bM_{\bx}^\dagger)\le \dim\mathcal T/\lambda\), and
\(\norm{\sum_i\bs_i a_i}^2\le\lambda^{-1}\sum_i a_i^2\) for \((a_1,\ldots,a_m)\in\R^m\).

Let \(b_i^F=[4f(\bx+c\bw_i/2)-f(\bx+c\bw_i)-3f(\bx)]/c-\ip{\nabla_{\mathcal T}f(\bx)}{\bw_i}\). The quadratic terms in the Taylor expansion of \eqref{eq:richardson} cancel, and
\(|b_i^F|\le L_2\norm{\bw_i}^3[4(c/2)^3+c^3]/(6c)\le C_1L_2R^3c^2\).
For \(\bw_i\ne\bzero\), independence of the two paired experiments and Assumption~\ref{ass:pair} imply
\begin{equation*}
\frac{16}{c^2}\Var[Y(\bx+c\bw_i/2,\xi_1)-Y(\bx,\xi_1')] +\frac{1}{c^2}\Var[Y(\bx+c\bw_i,\xi_2)-Y(\bx,\xi_2')] \le\sigma^2R^{2\kappa}(1+2^{4-2\kappa})c^{2\kappa-2}.
\end{equation*}
Let \(e_i^F=\widehat\mu_i^F-[4f(\bx+c\bw_i/2)-f(\bx+c\bw_i)-3f(\bx)]/c\). Unbiasedness and conditional independence imply \(\E[e_i^F]=0\), \(\E[e_i^Fe_j^F]=0\) for \(i\ne j\), and \(\E[(e_i^F)^2]\le\tau^2\lambda/\dim\mathcal T\). Hence,
\begin{equation}
\label{eq:reconstructionerror}
\E[\norm{\widetilde{\bg}-\nabla_{\mathcal T}f(\bx)}^2] =\big\|\sum\nolimits_i\bs_i b_i^F\big\|^2 +\sum\nolimits_i\norm{\bs_i}^2\E[(e_i^F)^2] \le C_1mL_2^2R^6c^4/\lambda+\tau^2.
\end{equation}

Write \(\bx^+=\Pi_{\mathcal X}(\bx-\eta\widetilde{\bg})\). For every \(\bx'\in\mathcal X\), the projection inequality is
\(\ip{\widetilde{\bg}}{\bx^+-\bx'}\le[\norm{\bx'-\bx}^2-\norm{\bx^+-\bx}^2-\norm{\bx'-\bx^+}^2]/(2\eta)\).
Combining this with smoothness, \(L_1\eta\le1\), and Young's inequality, we have
\begin{equation*}
\begin{aligned}
 f(\bx^+)
 &\le f(\bx)+\ip{\nabla_{\mathcal T}f(\bx)}{\bx'-\bx}
 +\norm{\bx'-\bx}^2/(2\eta)
 +\ip{\widetilde{\bg}-\nabla_{\mathcal T}f(\bx)}{\bx'-\bx^+}-\norm{\bx'-\bx^+}^2/(2\eta)\\
 &\le f(\bx)+\ip{\nabla_{\mathcal T}f(\bx)}{\bx'-\bx}
 +\norm{\bx'-\bx}^2/(2\eta)+\eta\norm{\widetilde{\bg}-\nabla_{\mathcal T}f(\bx)}^2/2.
\end{aligned}
\end{equation*}
Take \(\bx'=(1-\rho\eta)\bx+\rho\eta\bx_*\in\mathcal X\). Strong convexity implies
\(\ip{\nabla_{\mathcal T}f(\bx)}{\bx_*-\bx}\le f_*-f(\bx)-\rho\norm{\bx_*-\bx}^2/2\), so the squared-distance terms cancel and
\(f(\bx^+)-f_*\le(1-\rho\eta)(f(\bx)-f_*)+\eta\norm{\widetilde{\bg}-\nabla_{\mathcal T}f(\bx)}^2/2\).
Taking conditional expectations, first given the anchors and then given \(\cF_k\), and applying \eqref{eq:reconstructionerror} proves \eqref{eq:feasiblerecursion}. Also, \(\sup_{\bx\in\mathcal X}|f(\bx)|<\infty\) by compactness, so \(\E[V_k]<\infty\).

For possibly biased scalar estimates at tolerance \(\tau\sqrt{\lambda/m}\),
\(\E[\norm{\sum_i\bs_ie_i^F}^2]\le\lambda^{-1}\sum_i\E[(e_i^F)^2]\le\tau^2\).
Thus, \(\E[\norm{\widetilde{\bg}-\nabla_{\mathcal T}f(\bx)}^2]\le C_1mL_2^2R^6c^4/\lambda+2\tau^2\), and the projection bound establishes \eqref{eq:feasiblerecursion}.

For fixed anchors spanning the affine hull of \(\mathcal X\), let \(\overline{\bx}=m^{-1}\sum_i\bx^{[i]}\). For \(\mathbf v\in\mathcal T\),
\(\mathbf v^\top\bM_{\bx}\mathbf v=\sum_i\ip{\mathbf v}{\bx^{[i]}-\overline{\bx}}^2+m\ip{\mathbf v}{\overline{\bx}-\bx}^2\ge\sum_i\ip{\mathbf v}{\bx^{[i]}-\overline{\bx}}^2\).
Affine spanning implies \(\inf_{\bx\in\mathcal X}\lambda_{\min}(\bM_{\bx}|_{\mathcal T})\ge\min_{\mathbf v\in\mathcal T,\,\norm{\mathbf v}=1}\sum_i\ip{\mathbf v}{\bx^{[i]}-\overline{\bx}}^2>0\). For simplex vertices and \(\mathbf v\in\bone^\perp\),
\(\mathbf v^\top\bM_{\bx}\mathbf v=\sum_i(v_i-\bx^\top\mathbf v)^2=\norm{\mathbf v}^2+d(\bx^\top\mathbf v)^2\ge\norm{\mathbf v}^2\).\Halmos
\end{proof}

\begin{lemma}
\label{lem:propagation}
Suppose Assumptions~\ref{ass:objective} and \ref{ass:pair} hold. Let \(\eta_k\le1/(L_1d)\) for centered perturbations~\eqref{eq:qaceupdate} and \(\eta_k\le1/L_1\) for feasible reconstruction~\eqref{eq:feasibleupdate}. With the constants from Lemma~\ref{lem:onestep}, define
\begin{equation*}
 \psi_k:=
 \begin{cases}
  \mathcal C_2L_2^2c_k^4+\mathcal C_3L_1d^2\eta_k\tau_k^2,&\text{centered perturbations, unbiased},\\
  \mathcal C_2L_2^2c_k^4+\mathcal C_3d\tau_k^2,&\text{centered perturbations, possibly biased},\\
  \mathcal C_2mL_2^2R^6c_k^4/\lambda+\mathcal C_3\tau_k^2,&\text{feasible reconstruction}.
 \end{cases}
\end{equation*}
Then,
\begin{equation}
\label{eq:finitebound}
 \E[f(\bx_k)-f_*]\le\E[V_k]\le V_0\prod\nolimits_{j=0}^{k-1}(1-\rho\eta_j)
 +\sum\nolimits_{j=0}^{k-1}\eta_j\psi_j\prod\nolimits_{t=j+1}^{k-1}(1-\rho\eta_t).
\end{equation}
Moreover, \(\sum_{k=0}^{\infty}\eta_k=\infty\) and \(\sum_{k=0}^{\infty}\eta_k\psi_k<\infty\) imply \(\lim_{k\to\infty}\bx_k=\bx_*\) a.s.
\end{lemma}

\begin{proof}{Proof of Lemma~\ref{lem:propagation}.}
For centered perturbations, smoothness and \(\nabla f(\bx_*)=\bzero\) give \(f(\bx_k)-f_*\le L_1\norm{\bx_k-\bx_*}^2/2=V_k\); for feasible reconstruction, \(V_k=f(\bx_k)-f_*\). Iterating the expected recursion in Lemma~\ref{lem:onestep} proves \eqref{eq:finitebound}. With \(\cE_0\mathrel{:=}\E[f(\bx_0)-f_*]\), strong convexity gives \(V_0\le(L_1/\rho)\cE_0\) for centered perturbations, and \(V_0=\cE_0\) for feasible reconstruction.

For a.s. convergence, we apply Lemma~\ref{lem:onestep} to obtain 
\(
\E[V_{k+1}+\sum\nolimits_{j=k+1}^{\infty}\eta_j\psi_j\,|\,\cF_k]
 \le V_k+\sum\nolimits_{j=k}^{\infty}\eta_j\psi_j-\rho\eta_kV_k.
\)
The nonnegative supermartingale convergence theorem yields \(\lim_{k\to\infty}(V_k+\sum_{j=k}^{\infty}\eta_j\psi_j)\in[0,\infty)\) a.s. Summing the drift bounds and applying Tonelli's theorem, we obtain
\(\E[\sum_k\eta_kV_k]\le(V_0+\sum_k\eta_k\psi_k)/\rho<\infty\).
Since \(\lim_{k\to\infty}\sum_{j=k}^{\infty}\eta_j\psi_j=0\) and \(\sum_k\eta_k=\infty\), we obtain \(\lim_{k\to\infty}V_k=0\) a.s. Finally, note that \(\norm{\bx_k-\bx_*}^2\le2V_k/\rho\) for both updates. 
Hence, \(\lim_{k\to\infty}\bx_k=\bx_*\) a.s.\Halmos
\end{proof}

\begin{proposition}
\label{prop:querycost}
Under Assumptions~\ref{ass:objective} and \ref{ass:pair}, for centered perturbations~\eqref{eq:qaceupdate}, unbiased quantum mean estimation and classical paired averaging satisfy
\begin{equation}
\label{eq:queryq}
 \E[N_k^Q]=\cO\Big(\sum\nolimits_{j=0}^{k-1}\Big[1+\frac{2^{\kappa-1}\sigma}{c_j^{1-\kappa}\tau_j}\Big]\Big) \quad\text{and}\quad N_k^C=\cO\Big(\sum\nolimits_{j=0}^{k-1}\Big[1+\frac{2^{2\kappa-2}\sigma^2}{c_j^{2(1-\kappa)}\tau_j^2}\Big]\Big)\quad\text{a.s.}
\end{equation}
For feasible reconstruction~\eqref{eq:feasibleupdate} with unbiased estimates at scalar tolerance \(\tau_k\sqrt{\lambda/\dim\mathcal T}\),
\begin{equation*}
\begin{aligned}
 \E[N_k^Q]&=\cO\Big(m\sum\nolimits_{j=0}^{k-1}\Big[1+
 \frac{\sigma R^\kappa\sqrt{(1+2^{4-2\kappa})\dim\mathcal T/\lambda}}{c_j^{1-\kappa}\tau_j}\Big]\Big)\\
 \text{and}\quad N_k^C&=\cO\Big(m\sum\nolimits_{j=0}^{k-1}\Big[1+
 \frac{\sigma^2R^{2\kappa}(1+2^{4-2\kappa})\dim\mathcal T}{\lambda c_j^{2(1-\kappa)}\tau_j^2}\Big]\Big)\quad\text{a.s.}
\end{aligned}
\end{equation*}
For estimates with a deterministic query cap, the same quantum bounds hold a.s., with \(\dim\mathcal T\) replaced by \(m\) for feasible reconstruction~\eqref{eq:feasibleupdate} and scalar tolerance \(\tau_k\sqrt{\lambda/m}\).
\end{proposition}

\begin{proof}{Proof of Proposition~\ref{prop:querycost}.}
\label{app:general}
For the centered difference sampler, \(s_j/\tau_j=2^{\kappa-1}\sigma/(c_j^{1-\kappa}\tau_j)\). Lemma~\ref{lem:qvr} and the classical batch size \(\max\{1,\lceil s_j^2/\tau_j^2\rceil\}\), followed by summation over \(j<k\), prove \eqref{eq:queryq}. For feasible reconstruction~\eqref{eq:feasibleupdate}, substitute \(s^F\) at \(c=c_j\) and the stated scalar tolerances into the same bounds and sum over \(i\le m\) and \(j<k\).\Halmos
\end{proof}

\begin{proof}{Proof of Theorem~\ref{thm:diminishing}.}
\label{app:diminishing}
Define \(\cE_k\mathrel{:=}\E[f(\bx_k)-f_*]\).
The schedules imply \(\psi_k\le C_1(k+k_0)^{-\alpha}\) for some constant \(C_1>0\).
Choose \(C_2>0\) sufficiently large such that \(C_2\ge\max\{V_0k_0^\alpha,2C_1/\rho\}\). Then,
\(C_2[(k+k_0)^{-\alpha}-(k+k_0+1)^{-\alpha}]\le C_2\alpha(k+k_0)^{-\alpha-1}\le\bar\eta(\rho C_2-C_1)(k+k_0)^{-2\alpha}\).
Together with \(V_0\le C_2k_0^{-\alpha}\) and Lemma~\ref{lem:onestep}, this proves
\(\cE_k\le\E[V_k]\le C_2(k+k_0)^{-\alpha}\) by induction.
Since \(\sum_k\eta_k\psi_k\le C_1\bar\eta\sum_k(k+k_0)^{-2\alpha}<\infty\)
and \(\sum_k\eta_k=\infty\), Lemma~\ref{lem:propagation} implies
\(\lim_{k\to\infty}\bx_k=\bx_*\) a.s.; also,
\(\E[\norm{\bx_k-\bx_*}^2]\le2\E[V_k]/\rho=\cO((k+k_0)^{-\alpha})\).

By monotonicity, we have \(\sum_{j=0}^{k-1}(c_j^{1-\kappa}\tau_j)^{-p}\le k(c_{k-1}^{1-\kappa}\tau_{k-1})^{-p}\) for \(p\in\{1,2\}\). Hence, \eqref{eq:queryq} yields \(\E[N_k^Q]=\cO(k^{1+\alpha(1-\kappa)/4})\) and \(N_k^C=\cO(k^{1+\alpha(1-\kappa)/2})\) a.s. for QACE with centered perturbations. Together with \(\cE_k=\cO(k^{-\alpha})\), these bounds imply the stated rates under an expected budget.

For feasible reconstruction~\eqref{eq:feasibleupdate}, the additional factor \(\tau_k^{-1}=\cO(k^{\alpha/2})\) in Proposition~\ref{prop:querycost} yields the stated query orders.

For SPSA with one independent classical pair, \(\kappa=0\) and \(\tau_k=\sigma/(2c_k)\). Taking \(c_k=\bar c(k+k_0)^{-\alpha/6}\) yields \(\psi_k=\cO((k+k_0)^{-2\alpha/3})\) in \eqref{eq:qacerecursion}. The same comparison with exponent \(2\alpha/3\) yields \(\cE_k\le\E[V_k]=\cO((k+k_0)^{-2\alpha/3})\) and \(\sqrt{\E[\norm{\bx_k-\bx_*}^2]}=\cO((k+k_0)^{-\alpha/3})\).\Halmos
\end{proof}

\end{APPENDICES}

\bibliographystyle{informs2014}
\bibliography{references}
\end{document}